# Non-invasive digital etching of van der Waals semiconductors

Jian Zhou[1,2,†], Chunchen Zhang[1,3,†], Li Shi[4], Xiaoqing Chen[1,2], Tae-Soo Kim[5], Minseung Gyeon[5], Jian Chen[1,2], Jinlan Wang[4], Linwei Yu[1,2], Xinran Wang[1,2], Kibum Kang[5], Emanuele Orgiu[6], Paolo Samorì[7,*], Kenji Watanabe[8], Takashi Taniguchi[8], Kazuhito Tsukagoshi[8], Peng Wang[1,3,9,*], Yi Shi[1,2,*], Songlin Li[1,2,*]

[1] National Laboratory of Solid-State Microstructures and Collaborative Innovation Center of Advanced Microstructures, Nanjing University, Nanjing, China
[2] School of Electronic Science and Engineering, Nanjing University, Nanjing, China
[3] College of Engineering and Applied Sciences and Jiangsu Key Laboratory of Artificial Functional Materials, Nanjing University, Nanjing, China
[4] Department of Physics, Southeast University, Nanjing, China
[5] Department of Materials Science and Engineering, Korea Advanced Institute of Science and Technology, Republic of Korea
[6] Institut national de la recherche scientifique, Centre Énergie Matériaux Télécommunications, 1650 Blvd. Lionel-Boulet, J3X 1S2 Varennes, Canada
[7] University of Strasbourg, CNRS, ISIS UMR 7006, 8 allée Gaspard Monge, F-67000 Strasbourg, France
[8] National Institute for Materials Science, Tsukuba, Ibaraki 305-0044, Japan
[9] Department of Physics, University of Warwick, Coventry CV4 7AL, UK

[†] These authors contributed equally: Jian Zhou and Chunchen Zhang.
[*] Correspondence should be addressed to S.L. (sli@nju.edu.cn) or to P.S. (samori@unistra.fr) or to P.W. (wangpeng@nju.edu.cn) or to Y.S. (yshi@nju.edu.cn).

**Abstract.** The capability to finely tailor material thickness with simultaneous atomic precision and non-invasivity would be useful for constructing quantum platforms and post-Moore microelectronics. However, it remains challenging to attain synchronized controls over tailoring selectivity and precision. Here we report a protocol that allows for non-invasive and atomically digital etching of van der Waals transition-metal dichalcogenides through selective alloying via low-temperature thermal diffusion and subsequent wet etching. The mechanism of selective alloying between sacrifice metal atoms and defective or pristine dichalcogenides is analyzed with high-resolution scanning transmission electron microscopy. Also, the non-invasive nature and atomic level precision of our etching technique are corroborated by consistent spectral, crystallographic and electrical characterization measurements. The low-temperature charge mobility of as-etched $MoS_2$ reaches up to 1200 $cm^2V^{-1}s^{-1}$, comparable to that of exfoliated pristine counterparts. The entire protocol represents a highly precise and non-invasive tailoring route for material manipulation.

Modern science and technology have benefited vastly from the ever-increasing capability of fine control on material dimensions. For instance, in two-dimensional (2D) van der Waals materials, the reduced dimensionality by approaching the atomic thickness can result in major changes in fundamental physical characteristics such as the density of states, band structures[1,2], crystal electrostatic fields[3,4] and even lattice symmetries, which allows for emerged couplings of material parameters, such as electron wavefunction radii and magnetic interaction lengths. This provides a platform for exploring intriguing physical phenomena such as Dirac fermions[2], valley chirality[5], Moiré superlattices[3,4], and thickness sensitive magnetism[6,7]. On the other hand, the continuous downscaling in modern microelectronics has brought the industry to technology nodes of several nanometers[8]. In the near future, a finer control up to atomic levels will be soon required[9]. In this context, atomically thin 2D semiconductors, in particular monolayer transition-metal dichalcogenides (TMDCs) which exhibit ideal surface flatness and sizeable carrier mobility, are regarded as promising channel materials in the post-silicon era[9-11]. For achieving large-area atomic channels, the most straightforward but yet challenging route is to bottom-up grow homogeneous monolayer wafers directly, while it also represents a possible route to tailor the pristinely inhomogeneous few-layer wafers into homogeneous ones through selective top-down etching.

For both the fundamental and applied purposes, it is highly anticipated that the original crystallinity, thus the intrinsic physical properties of the materials, are preserved after material processing. In this regard, a protocol enabling simultaneously non-invasive and digital etching of TMDCs, that is precise layer-by-layer removal of the topmost monolayer while keeping the underlying intact, represents a keen anticipation and, on the same time, a big challenge. Previous approaches developed for atomic etching of van der Waals TMDCs, including plasma bombardment[12–14], laser treatment[15,16], and thermal oxidation[17], are all proven invasive to materials because they generally exert less selective etching effects on the layers to be removed and preserved, thus resulting in degraded electronic performance. To date, the techniques that allow for non-invasive etching of the 2D van der Waals crystals are still highly sought after.

Here we report a soft protocol for non-invasive digital etching of van der Waals TMDCs by exploiting the selective thermal diffusion of a sacrificial metal into assigned TMDC layers and subsequent removal of the surfacial alloy layer. Synergic strategies including surface defect engineering and low-temperature annealing were devised to strictly limit the etching depth to a monolayer at each etching cycle. The nature of layer-by-layer digital etching was confirmed by aberration-corrected high-resolution scanning transmission electron microscopy (HR-STEM), atomic force microscope (AFM), secondary harmonic generation (SHG) and Raman spectroscopy. Both the crystallographic and electrical characterizations suggest that the high lattice quality and intrinsic transport performance are preserved in the as-etched samples, implying the non-invasivity of the protocol. Besides the non-invasivity and atomic-level precision, the strategy also features important advantages including multiple processability, universality for different TMDC materials and CMOS compatibility. The underlying physics for the selective thermal diffusion between nearly perfect pristine and defective van der Waals lattices is also extensively studied. The research results well address a grand etching challenge confronted in material science and open a door to thoroughly exploit the van der Waals TMDC crystals for quantum and electronic purposes.

**Results and Discussion**

**Rationale for defect engineering.** Our protocol for the non-invasive digital layer-by-layer etching mainly comprises the following steps. First, metal Al sacrificial layers are deposited selectively onto local defect-engineered TMDC areas to be thinned. Then, controlled thermal diffusion of Al into TMDCs is achieved by tuning the temperature and duration in the subsequent step of thermal annealing. Finally, the layers of Al and its alloy with TMDCs are wet etched with acid or basic solutions. We emphasize that, without appropriate defect engineering in the first step, a simple implementation of such protocol could cause uncontrollable Al diffusion and thus random etching into underlying layers in our preliminary attempts, as depicted in Fig. 1a. In circumstance that no artificial defects are produced, we found that the thermal diffusion of Al into TMDCs still relies on the pristine, randomly distributed vacancies in lattices[18], which result in disordered diffusion inside the van der Waals materials, because they can reduce local diffusion energy barriers and serve as the starting sites to guide the disordered diffusion, as shown in Fig. 1c.

**Uncontrollable thermal diffusion of Al into pristine MoS₂.** In the preliminary experiments where no artificial defect engineering was performed, we first investigated the thermal diffusion mechanism of Al into freshly mechanically exfoliated $MoS_2$, which is proven to be the defect-assisted diffusion proess as mentioned above, by simply exploring the dependence of diffusion rate on annealing temperature ($T_a$). Figure 1b–d shows the optical images of three $MoS_2$ samples washed with acid, after one-hour thermal diffusion of Al at $T_a$ = 250, 300 and 350 °C, respectively. With increasing $T_a$ the corrosion pits become deeper and spatially more extended, as indicated by the dashed rectangles. Indubitably, $T_a$ is an important parameter to promote the Al atoms to overcome the diffusion barriers within $MoS_2$ and, normally following the Arrhenius relationship, to determine the diffusion rate and overall etching depth[18]. Noteworthy, at low $T_a$ (up to 250 °C) no appreciable traces of Al diffusion and corrosion pits are observed, suggesting a negligible diffusion of Al into the pristine (nearly perfect) $MoS_2$ lattices. For samples annealed at intermediate $T_a$ (300 °C), a mixture of corrosion pits and unetched islands are monitored in Fig. 1c, indicating that the thermal diffusion of Al into pristine $MoS_2$ lattice is a poorly uniform process, mainly guided by the disordered lattice vacancies present in pristine samples. Although the pristine $MoS_2$ can be thinned at this temperature, the thickness reduction features a modest control.

Cross-sectional high-angle annular dark-field imaging (HAADF) and energy-dispersive X-ray spectroscopy (EDS) by HR-STEM made it possible to monitor the boundaries between layers to further unveil the interlayer diffusion behavior at different $T_a$s. Figure 1e–i shows the typical HR-STEM images and corresponding EDS elemental mappings for the $Al/MoS_2/SiO_2$ stacks annealed at 250 °C. They exhibit two sharp interfaces for the encapsulated $MoS_2$ layers, validating the negligible diffusion of the upper Al or the underlying $SiO_2$ into the encapsulated $MoS_2$ layers at such a low $T_a$. Conversely, the images for the stack annealed at 300 °C are displayed in Fig. 1j–m.

The HAADF image displayed in Fig. 1j is informative. First, it shows that the interlayer diffusion of Al and $MoS_2$ indeed occurs at this elevated temperature, resulting in the delamination of local $MoS_2$ up to 6 layers in the central area. The $Al/MoS_2$ alloy area indicated by dotted lines resembles a surface bubble. Second, the thermal diffusion of metal atoms into the van der Waals materials appears as a rather ordered process along the vertical direction that tends to terminate at the van der Waals gaps and gives rise to a clear diffusion boundary in the central area. Thus, it is expected that the breadth of the diffusion depth can be further finely controlled within a monolayer at a low $T_a$ below 300 °C. Third, a mixed combination of alloy and unreacted areas (sharp boundaries), with roughly equal probabilities, are observed in the extended HR-STEM images (Supplementary Fig. 2a), suggesting the diffusion mechanism in pristine $MoS_2$ as follows. Given the low energy of formation of sulfur vacancies and the trend of sulfur loss at elevated temperatures[19-21], it is thus deduced that the random Al diffusion is associated with the disordered sulfur vacancies pristinely present and/or thermally created. They reduce the diffusion barriers of local areas and act as permeation paths for external atoms, resulting in the disordered Al diffusion through the underlying $MoS_2$ layers[18,22].

**Principle of controllable diffusion via defect engineering.** To circumvent the issue of random and excessive Al diffusion, we devised a "selective diffusion" strategy aimed at producing uniform diffusion sites and, at the same time, spatially confining diffusion depth within the thickness of one monolayer. The basic idea is to predefine surfacial lattice vacancies distributed uniformly in the topmost layer before thermal diffusion, which is expected to reduce only the diffusion barrier of the topmost layer and produce a vast difference in the diffusion rates, which increases exponentially by lowing diffusion barrier, of Al atoms between the defective topmost and pristine underlying TMDC layers. Theoretical studies revealed that Ar plasma irradiation can produce various single- and multiple-atom vacancies (e.g., S, Mo and $MoS_6$) into $MoS_2$ lattices[23]. Hence, by applying a controlled short duration of Ar plasma irradiation and introducing a limited low density of vacancies in the surfaces, the topmost layer can be tailored to exhibit much larger diffusion rates than the underlying layers. Accordingly, the topmost and underlying layers would exhibit selective overall etching rates, which is beneficial for selectively etching the topmost layer while minimizing the negative corrosive impact on the underling layers. In addition, we try to minimize $T_a$ as low as 250 °C, which can ensure regular diffusion rates for Al into defective topmost layer but negligible rates into pristine underlying layers, as proven in Fig. 1b.

Theoretical calculations also indicate that the diffusion barrier for Al via sulfur vacancies into a defective $MoS_2$ lattice is about 90 meV, being 8 times lower than in a pristine crystal (Supplementary Note 5). According to the Fick's law of diffusion[18], the 8-fold variation in diffusion barrier corresponds to a huge difference of $10^6$ in thermal diffusion coefficients between the defective and pristine lattices at 250 °C, constituting the rationale behind the selective etching.

**Layer-by-layer digital etching.** With the "selective diffusion" strategy in mind, we then perform the devised non-invasive digital etching protocol on defective engineered $MoS_2$ samples. Figure 2a–d illustrates the processing flow and corresponding optical images at each step for the soft etching technique. Initially, a low-energy Ar plasma irradiation is applied onto exfoliated $MoS_2$ layers to introduce vacancies into the topmost $MoS_2$ layers (Fig. 2a). The irradiation energy and duration are carefully controlled to ensure that only the topmost sulfur atoms are affected without damaging the underlying layers (Supplementary Notes 8 and 9). Then, a 10 nm metal Al strip is deposited onto $MoS_2$ (Fig. 2b), followed by a diffusion process at 250 °C for 0.5 h (Fig. 2c). At such low $T_a$, the Al atoms diffuse mainly through the defective topmost layer, leaving the underlying layers almost intact. Lastly, the Al strip and its alloy with the topmost $MoS_2$ layer are dissolved in hydrochloric acid and, consequently, a monolayer of $MoS_2$ is removed (Fig. 2d).

A critical requirement for the digital etching technique is to precisely control the etching depth down to one monolayer. To check the etching depth and surface flatness of the as-etched area, we performed AFM measurements on the as-etched samples in order to determine the thickness change and surface topography. As shown in Fig. 2e, after one-cycle (1C) processing, the step height between the pristine (0C) and etched areas is about 0.67 nm, which is consistent with the theoretical value of the thickness of a $MoS_2$ monolayer, being 0.615 nm[24], and verifies the nature of layer-by-layer etching of this technique. Besides, the root mean square roughness of the etched area is estimated to be 0.38 nm, which is comparable to that of the pristine area ~0.43 nm, proving the excellent surface quality of the as-etched area. Cross-sectional and top-view images recorded by HR-STEM also confirmed the nature of layer-by-layer tailoring of this technique. In Fig. 2f a sharp monolayer step produced by etching can be clearly seen in the cross-sectional image taken at the trilayer/tetralayer (3L/4L) step while in Fig. 2g a clear contrast between the monolayer (1L) and bilayer (2L) areas can be observed in the top-view image taken for an etched 2L sample. Moreover, the HR-STEM images can be used to accurately evaluate the lateral resolution of our etching technique, owing to unwanted lateral diffusion. After analysis on the edge profiles of the alloy areas, we conclude that the lateral resolution at above etching condition is better than 1.5 ± 0.3 nm. Detailed discussion can be found in Supplementary Note 4.

To further confirm the nature of digital layer-by-layer etching, Raman spectra were also recorded for a twice-etched sample, which contains 0C, 1C and 2C etched areas, as shown in Fig. 2h. It is well known that the thickness of few-layer TMDCs can be determined from the distance of their characteristic Raman modes[25,26]. With the decrease in MoS$_2$ thickness, the distance between the $E_{2g}^1$ and $A_{1g}$ modes is decreased accordingly. As shown in the inset of Fig. 2h, the distance values for the three contrastive areas amount to 23.6 cm$^{-1}$, 22.2 cm$^{-1}$ and 19 cm$^{-1}$, which correspond to the thickness values of 3L, 2L and 1L, respectively. These consecutive numbers of thickness provide unambiguous evidence for the accurate thickness control featuring monolayer precision and the way as digital tailoring in the process.

Since this etching method finds its roots in the diffusion and removal of Al atoms, which may reside on the surfaces of the as-etched materials, it is also important to check the lattice quality and trace of Al residues atop the materials. To cast light onto this issue, we also performed atomically resolved top-view STEM and EDS elemental analyses for as-etched MoS$_2$ monolayers, as shown in Supplementary Fig. 12 and Fig. 2i–l. In a series of atomic STEM images, we performed a statistical analysis to estimate the gross vacancy density of about (1.3±0.6)×10$^{13}$ cm$^{-2}$, without ruling out the extra bombardment effect from the electron irradiation during STEM imaging. This value is comparable to (1.2±0.4)×10$^{13}$ cm$^{-2}$, the value reported in high-quality exfoliated sheets[27], indicating the negligible invasiveness of our etching technique. In Fig. 2i, the EDS spectrum collected over large-area reveals that the trace of Al residues, located at 1.49 keV, is below the instrumental uncertainty of 5%. Elemental mappings for Al, Mo and S also indicate there is no detectable Al signals (Fig. 2j), as compared with the strong Mo and S signals (Fig. 2k, l). Note that the sparse bright pixels in Fig. 2j arise likely from the "background" chamber contamination and random noise of the imaging system. Detailed discussion can be found in Supplementary Fig. 13. Besides, XPS analysis (Supplementary Fig. 14) also supports the conclusion that the trace of Al residues is negligible.

**Alternative method for defect engineering.** As far as surface defect engineering is concerned, besides the Ar plasma irradiation, there are also various methods to introduce vacancies into the topmost layer, such as thermal decomposition. We tested this method by pre-annealing for 1 h the freshly exfoliated MoS$_2$ samples at 400 °C, a temperature near the critical point for the escape of sulfurs on the surface, before Al deposition. Under such a pre-annealing condition, the density of sulfur vacancies is greatly enhanced at the MoS$_2$ surface, although the surface exhibits no remarkable change under optical microscope (Supplementary Fig. 15a). As a result, after the low-temperature annealing together with a sacrificial Al strip at 250 °C for 0.5 h (Supplementary Fig. 15b) and acid wash, the topmost layer is clearly removed, as shown in Supplementary Fig. 15c. We note that this process would not take place without the artificial defects introduced by proper pre-annealing. Hence, the thermal decomposition of the surface sulfur atoms via pre-annealing also represents an appropriate defect engineering strategy that result in the layer-by-layer etching, as it follows the concept of selective etching.

**Multiple digital etching.** For both lab and industrial applications, complex patterning and multiple etching steps are highly sought after. Next, we demonstrated the possibility of applying this procedure repeatedly to construct complex patterns when combined with electron beam lithography (EBL). Figure 3a shows the optical image of a checkerboard-like pattern on a MoS$_2$ sheet that is defined by alternatively applying the etching procedures (defect engineering via Ar plasma irradiation) along the vertical and horizontal directions. According to such patterning motifs, local areas with the three consecutive numbers of layers n, n-1, and n-2 are constructed, where n is the numbers of unetched layers. The success in this patterning characterized by consecutive numbers of layers, can be clearly seen by means of optical, AFM, and Raman characterizations. According to previous research[25], the ratio of the Raman modes of MoS$_2$ to Si substrate can be used to discern samples characterized by a few stacked TMDC layers with number of layers up to 10. In Fig. 3b, c, we mapped the area ratio of the Raman modes of $E^1_{2g}$ and $A_{1g}$ to Si, respectively, where the checkerboard-like pattern is clearly seen.

Second-harmonic generation (SHG) response[28], being a sensitive probe for the lattice symmetry of thin TMDCs, was also employed to verify the variation of number of layers. It has been shown that MoS$_2$ sheets with odd numbers of layers display a significant SHG response while those with even numbers of layers exhibit almost zero response. Figure 3d displays the SHG mapping for the sample; it exhibits alternative light and dark patterns along the vertical and horizontal directions, in good accordance with the checkerboard-like pattern. All these characterizations indicate that our soft etching technique is atomically precise in achieving uniform layer-by-layer etching, with no obvious residues after each etching step.

We also exploited this method to various TMDCs with more delicate patterns to demonstrate its universal applicability. Figure 3e–g portrays the optical images of patterned logos of Nanjing University on three different TMDCs: MoS$_2$, WS$_2$ and WSe$_2$, respectively. The nearly identical logo patterns imply that the concept of selective etching is likely applicable for various TMDCs and van der Waals materials, if they can be treated in appropriate conditions to ensure required defect density in topmost layers, temperature and duration of thermal diffusion, concentration of wash solution, etc. Raman and photoluminescent spectra were also employed to characterize pristine and etched WS$_2$ and WSe$_2$ sheets, as shown in Supplementary Fig. 16. All data support the validity of our method.

**Electronic evidence for non-invasive etching.** In order to evaluate the effect of extra lattice defects and Al residues on the overall electrical quality, cryogenic electrical characterizations were carried out on as-etched MoS$_2$ layers. After etching, the remained MoS$_2$ samples were transferred and encapsulated by two ultraclean hexagonal boron nitride (h-BN) dielectrics and standard Hall geometry was adopted to estimate its intrinsic electrical performance, as shown in the inset of Fig. 4a. Figure 4a shows typical transfer curves (four-terminal conductivity $\sigma$ versus $V_g$) for an FET channel made from as-etched 4L MoS$_2$ at different temperature ($T$) vales from 10 to 300 K. The device exhibits high current on/off ratios of $10^9-10^{11}$, suggesting that the semiconducting nature is well preserved in the as-etched sample. The curves intersect around $V_g \sim 0$, indicating the emergence of metal-insulator transition upon modulating carrier concentration. This behavior is consistent with that reported in high-quality samples[29] and confirms again the preservation of high crystallinity after etching.

In Fig. 4b, we plot intrinsic carrier mobility ($\mu$) versus $T$ to further discern the effect of carrier scattering from Al residues. At low $T$, $\mu$ becomes saturated at ~1200 cm$^2$V$^{-1}$s$^{-1}$. This value is consistent with exfoliated samples[29]. At high $T$, $\mu$ follows a power law with $T$ ($\mu \propto T^{-\gamma}$ with $\gamma \sim 1.83$) in the log-log plot. Theoretical studies on pure phonon scattering predicts an exponent $\gamma \sim 1.69$ and ~2.6 for 1L and bulk MoS$_2$, respectively[30]. However, the addition of Coulomb impurity scattering will degrade $\mu$ in the entire $T$ regime and, as a result, reduce $\gamma$ largely. For instance, $\gamma \sim 0.62-0.72$ were reported in SiO$_2$ supported monolayer MoS$_2$ samples[31,32]. The $\gamma$ value in our device is reasonably within the range of theoretical prediction, suggesting the dominancy of scattering from lattice phonons and the negligibility of effect from Coulomb impurities such as Al residues. The entire $\mu - T$ trend of the as-etched samples is comparable to that of h-BN encapsulated exfoliated counterparts[29], thus it can be inferred that the acid wash can remove nearly all the Al residues and provide a clean and fresh surface for the as-etched samples.

Finally, we analyze the transport mechanism of the etched layer based on the variable $T$ transport characteristics. As previously reported[31,33], due to the atomic thickness, the surface disorders, such as the adsorbates atop and substrate charges underneath, tend to cause Anderson localization, especially at low carrier densities. Even in the case of highly crystalline materials, the presence of a high density of localized states in the band-gap region can lead to variable-range hopping carrier transport behavior in few-layer MoS$_2$. To gain further insight into the transport mechanism, we plot in Fig. 4c $\sigma$ versus $T$ in terms of the modified 2D Mott variable-range hopping (VRH) equation[31,33]

$$\sigma = A \cdot T^m \cdot \exp(-\left(\frac{T_0}{T}\right)^{1/3}), \qquad (1)$$

where $A$ is a constant, $m$ is the coefficient normally adopted as 0.8 for 2D TMDCs, and $T_0$ is the characteristic temperature. The parameter $T_0$ is related to the localization length $\xi(E)$ which can be used to qualitatively estimate the degree of disorder, and can be described as

$$T_0 = \frac{13.8}{k_B \xi^2(E) D(E)}, \qquad (2)$$

where $k_B$ is the Boltzmann constant, and $D(E)$ is the typical density of trap states from interfacial residues. The fitted values of $T_0$ is plotted versus $V_g$ in the inset of Fig. 4d. $T_0$ keeps fixed around 300 K, which is 1-3 orders lower in magnitude than the values reported in literature[31,33], indicating the large localization length and insignificant disorders induced by Al residues. This feature corroborates further the insignificance of the extrinsic disorders introduced by Al residues.

In summary, we have devised a selective etching protocol enabling the atomically precise and non-invasive layer-by-layer etching of 2D van der Waals materials, based on thermal diffusion and subsequent alloy dissolution of a local metal sacrificial layer. Such an etching protocol addresses the challenge of non-invasive tailoring material thickness with an atomic precision. This technique is universal, being applicable to different van der Waals TMDC materials, as it only requires the van der Waals materials to be stable in acid or basic solutions. The realization of atomic precision in the etching process and the high crystalline quality of the as-etched layers (i.e., surface morphology and electronic transport) is demonstrated by a series of characterization means including HR-STEM, AFM, Raman, SHG response, and $T$ variable electronic measurement. Our protocol for *in situ* layer-by-layer etching is easily up-scalable, thus suitable for both lab and industrial applications.

**Methods**

**Pristine crystals.** The MoS$_2$ crystal was selected from natural minerals, WS$_2$ purchased from 2D Semiconductors Inc., and WSe$_2$ synthesized by chemical vapor transport. All thin TMDC layers used in experiment to be etched were mechanically exfoliated from corresponding crystals and transferred to Si substrates with a 90 or 285 nm $SiO_2$ capping layer.

**Etching procedure.** At first, the exfoliated TMDC layers were spin coated with a layer of A4 PMMA electron resist and the areas to be etched were then exposed by EBL, transferred to an inductively coupled plasma (ICP) chamber for Ar plasma irradiation. The ICP chamber was initially pumped to a base vacuum of $8\times10^{-3}$ Pa and was then filled with a constant Ar flow of 10 sccm till the working pressure of 0.5 Pa. The TMDC samples were then irradiated in Ar plasma at an ICP generator power of 30 W and an idle CCP biasing power for 30 s. The CCP power was set in the idle mode in all experiment to minimize plasma energy and the attacking depth. The energy of Ar plasma is about 60–70 eV. Subsequently, the Al sacrifice strips were deposited on the TMDC samples by thermal evaporation at a rate of 0.7–1.2 Å/s. A thickness of 10 nm was employed to ensure complete coverage of the areas to be etched. After liftoff, the TMDC samples partially covered with the Al strips were then annealed under a 2 torr nitrogen atmosphere. The annealing recipe for the layer-by-layer etching is 250 °C for 0.5 h, while the recipes for the control experiments are 250, 300 and 350 °C for 1 h to determine the diffusion depths of Al into $MoS_2$ at different temperatures. Finally, the annealed samples were immersed into dilute hydrochloric acid (~16 %) for 0.5 h, to completely remove the unreacted Al and its alloy with TMDCs, and then were rinsed in deionized water multiple times.

**Raman and SHG mapping.** Raman spectra were acquired at an excitation wavelength of 488 nm with a power of 22 mW. The integration time was 5 s when collecting spectra for individual points and was reduced to 1 s in the mapping mode to avoid loss of focus during the long-time collection. SHG was excited by a 1550 nm fiber laser with an 80 MHz repetition rate. The laser power is set at 100 mW and the integration time used is 100 ms.

**HR-STEM imaging.** The cross-sectional STEM samples were fabricated by a lift-out method using focused ion beam technique (FEI Helios 600i dual-beam system). The STEM lamellas were thinned down to a thickness below 100 nm using a beam current of 0.79 nA at 30 kV, followed by gentle milling with a beam current of 72 pA at 2 kV. The top-view STEM samples for atomic imaging were prepared by direct transfer of as-etched $MoS_2$ monolayers from Si substrates onto 300-mesh copper STEM grids through polymer PMMA as transfer medium. HR-STEM HAADF images and EDS mapping were acquired on a double aberration-corrected FEI Titan Cubed G2 60-300 S/TEM at 60 kV equipped with a Super-X EDS detector.

**Device fabrication and characterization.** The mechanically exfoliated $MoS_2$ samples were firstly transferred by Poly-(propylene carbonate) (PPC) films from $Si/SiO_2$ substrates onto ultraclean h-BN flakes exfoliated on other $Si/SiO_2$ substrates, followed by our non-invasive etching method to remove the top layer. Then, few-layer graphene Hall electrodes that were pre-etched were picked-up by top h-BN flakes supported with PPC/PDMS bilayers. Subsequently, the PPC layers were heated till softened to release the h-BN/graphene bilayers as the top encapsulator/electrode onto the as-etched $MoS_2$/bottom h-BN bilayers to form the target h-BN/graphene/$MoS_2$/h-BN structures. Afterwards, electrode vias were selectively opened on the top h-BN layers by standard EBL and $CF_4$ plasma exposure to expose the graphene Hall electrodes. Finally, steps of EBL and metallization of 10 nm Ni/50 nm Au were carried out to wire out the graphene Hall electrodes for electrical characterization. Electrical measurements were performed in vacuum at $10^{-5}$ Torr using a probe station (CRX-6.5K, Lake Shore) and two Keithley 2636B sourcemeters. The values of four-terminal field-effect mobility of as-etched $MoS_2$ channels were calculated by using the equation $\mu = \frac{1}{C_{\text{ox}}} \cdot \frac{d\sigma}{dV_{\text{g}}}$, where the $\sigma$ is the normalized conductance of the channels, $C_{\text{ox}}$ is the capacitance of the bilayer dielectric structure of h-BN (~10 nm) and SiO$_2$ (90 or 285 nm).

**Data availability**
The data that support the plots within this paper and other finding of this study are available from the corresponding authors on reasonable request.

**Acknowledgments**

The activity in Nanjing was supported by the National Key R&D Program of China (2017YFA0206304), the National Natural Science Foundation of China (61974060, 61674080, 61521001 and 11874199), the Fundamental Research Funds for the Central Universities (021014902310, 021014380116 and 020514380224), the Innovation and Entrepreneurship Program of Jiangsu province and the Micro Fabrication and Integration Technology Center in Nanjing University. The activity in Strasbourg was supported by the EC through the ERC project SUPRA2DMAT (GA-833707) and the Graphene Flagship Core 3 project (GA-881603) as well as the Labex projects CSC (ANR-10LABX-0026 CSC) and NIE (ANR-11-LABX-0058 NIE) within the Investissement d'Avenir program ANR-10-IDEX-0002-02 and the International Center for Frontier Research in Chemistry. K.W. and T.T. acknowledge support from the Elemental Strategy Initiative conducted by the MEXT, Japan (Grant Number JPMXP0112101001) and JSPS KAKENHI (Grant Numbers 19H05790, 20H00354 and 21H05233). KK acknowledges the support from the National Research Foundation of Korea (2020M3F3A2A01082618 and 2020M3F3A2A01081899).

**Author contributions**
S.L, and J.Z. conceived the experiments. J.Z. performed the thermal diffusion, AFM, Raman, device fabrication and electrical measurements. C.Z. and P.W. performed the HR-STEM characterization and analyses. L.S and J.W. performed the theoretical calculations. J.C. and L.Y. performed the EBL and plasma etching. X.C., and X.W. performed the SHG mapping. K.W and T.T. provided the h-BN crystals. T.K., M.G. and K.K. provided the MOCVD $MoS_2$ as control samples. S.L., J.Z., E.O., K.T. and Y.S. discussed the results and contributed to the interpretation of data. S.L., J.Z., and P.S. co-wrote the paper with inputs from other co-authors. All authors commented on the manuscript.

**Competing interests**
The authors declare no competing interests.

**Figure captions**
**Fig. 1 Uncontrollable thermal diffusion of Al atoms into pristine $MoS_2$ lattices. a,** Schematic diagram for the thermal diffusion process, where randomly distributed surficial defects serve as the starting sites to guide the diffusion inside the $MoS_2$ lattices. **b-d,** Typical optical images of three acid-washed $MoS_2$ samples after one-hour thermal diffusion of Al at $T_a$ = 250, 300 and 350 °C, respectively. The dashed yellow rectangles outline the areas of locally deposited Al sacrifice metal. Scale bars, 5 μm. **e-i,** Cross-sectional HAADF image and corresponding elemental mappings for a typical Al/$MoS_2$/$SiO_2$ stack annealed at 250 °C. The dashed white lines in **f-i** highlight the two sharp interfaces of $MoS_2$. Scale bars in **e** and **f-i** are 20 and 6 nm, respectively. **j-m,** Cross-sectional HAADF image and corresponding elemental mappings for a typical Al/$MoS_2$ stack annealed at 300 °C. The dotted white line in **j** shows the alloy region between Al and $MoS_2$. The dashed white lines in **k-m** highlights the interfaces of Al/alloy and alloy/$MoS_2$. Scale bars in **j** and **k-m** are 40 and 6 nm, respectively.

**Fig. 2 Non-invasive digital etching technique and corresponding characterization. a-d,** Schematic processing flow for the controllable digital etching. Scale bar: 10 μm. **a,** Controlled Ar plasma irradiation to produce uniformly distributed sulfur vacancies in the topmost $MoS_2$ layer. **b,** Local deposition of sacrificial metal Al strips on $MoS_2$. **c,** Thermal annealing at 250 °C for 0.5 h to facilitate the diffusion of Al into $MoS_2$. **d,** Dissolving Al and related alloy by hydrochloric acid. The topmost $MoS_2$ layer is removed finally, where nL denotes the numbers of layers. Digitally etched monolayer steps (0.67 nm) between the pristine and etched areas as revealed by **e** AFM, **f** cross-sectional and **g** top-view atomic images. Scale bar in **e**: 3 μm. Scale bar in **f**, **g**: 2 nm. **h,** Raman spectra for pristine (0C) and etched local $MoS_2$ areas for one cycle (1C) and two cycles (2C). Inset: Peak distance between the $E^1_{2g}$ and $A_{1g}$ modes versus etching cycles (i.e., number of $MoS_2$ layers). **i,** Accumulated energy dispersive spectrum from the as-etched monolayer to estimate the content of Al residues. **j-l,** Typical EDX elemental mappings for the Al, Mo and S elements, respectively. The trace of Al residues is below the instrumental uncertainty of 5%. Scale bars in **j-l**: 1 nm.

**Fig. 3 Fabricating complex patterns on different TMDCs. a,** A checkerboard-like motif on $MoS_2$ defined with two stripy etching cycles along the vertical and horizontal directions. Scale bar: 6 μm. Inset: the AFM image scanned from a local area of the checkerboard-like motif. The label n, n-1 and n-2 represent the numbers of local $MoS_2$

Published at *Nature Communications* 13, 1844 (2022), doi: 10.1038/s41467-022-29447-6areas. Scale bar, 3 μm. **b-d,** Raman and SHG mapping for the local area denoted by the red dashed rectangle in **a**. Raman mappings were acquired by calculating the area ratio of the **(b)** $E^1_{2g}$ and **(c)** $A_{1g}$ modes of MoS$_2$ to the 520 cm$^{-1}$ mode of Si. The bright and blue areas in the SHG mapping in panel **d** denote the local areas with odd and even number of MoS$_2$ layers. Scale bars in **b-d**, 3 μm. **e-g,** The patterned logos for Nanjing University on exfoliated MoS$_2$, WS$_2$ and WSe$_2$, respectively. Scale bars, 6 μm.

**Fig. 4 Electrical measurement for a typical as-etched 4L MoS$_2$ encapsulated by ultraclean h-BN dielectrics. a,** Transfer characteristics for the as-etched 4L MoS$_2$ at various different *T* values of 10 K, 30 K, 50 K, 80 K, 120 K, 180 K, 240 K and 300 K. Inset: Optical image for the BN/Graphene/MoS$_2$/BN structure where graphene (labelled by dashed blue lines) is used as the electrodes for MoS$_2$ channel (dashed red lines) in the standard Hall geometry. Scale bar, 4 μm. **b**, *T*-dependent field-effect mobility. At *T* > 60 K, the mobility follows the power law $\mu \propto T^{-\gamma}$ with *γ* = 1.83. The black dashed line ($\sim T^{-1.83}$) is a guide to the eyes. Inset: $\mu - T$ curve of a typical mechanically exfoliated multi-layer MoS$_2$. Adapted with permission from Ref. 29. Copyright 2015 American Chemical Society. Our etched sample exhibits comparable performance to exfoliated counterparts. **c,** *T*-dependent normalized conductance at various gate voltages. At low *T* regime the carrier transport can be described by Mott VRH model, while it turns into the band-like transport behaviour at high *T* regime. **d,** Values of characteristic temperature ($T_0$) estimated from the Mott VRH model at low *T* regime versus gate voltage. The values are rather small, indicating that the disorder due to Al residues is insignificant.

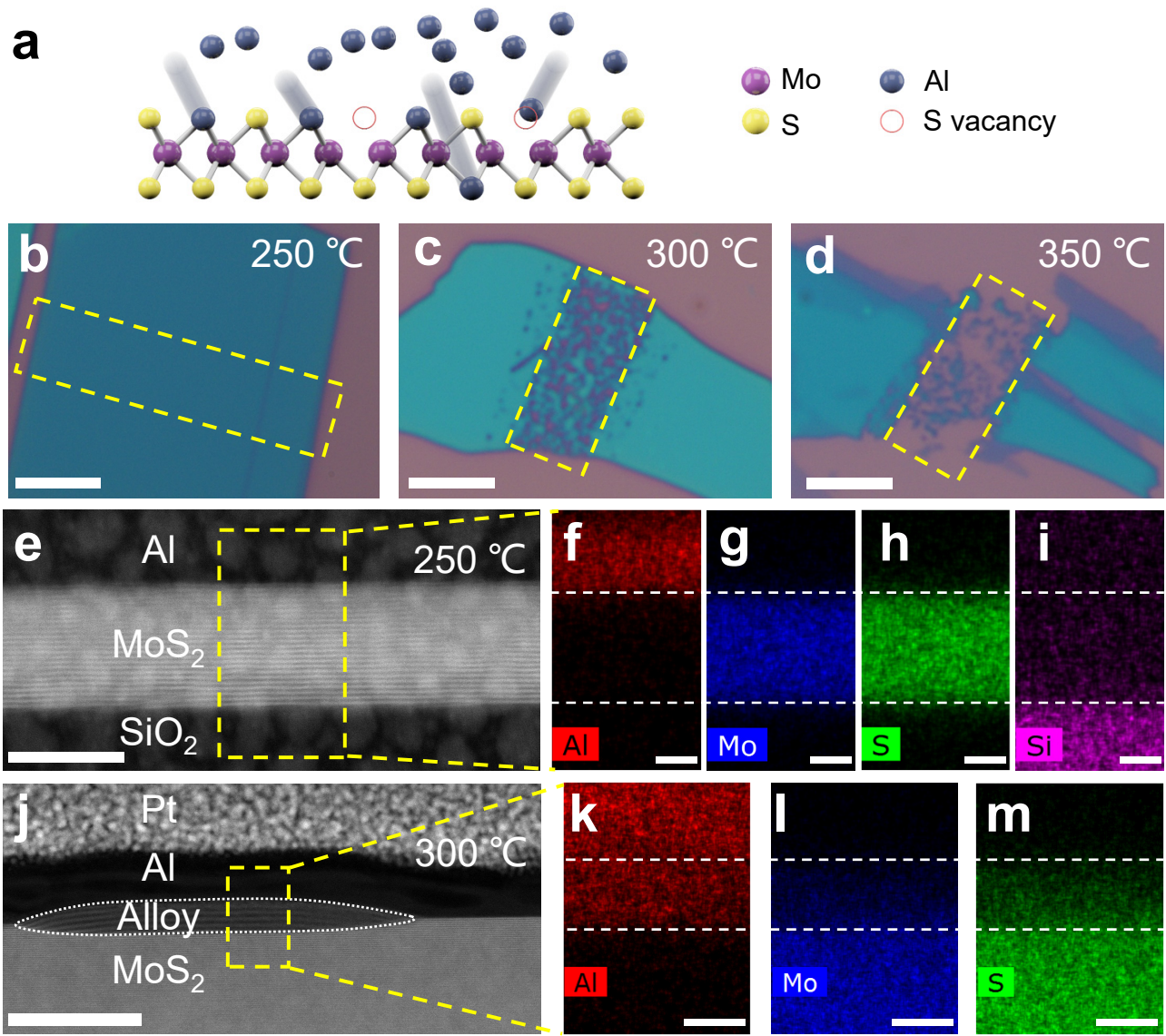

Fig.1

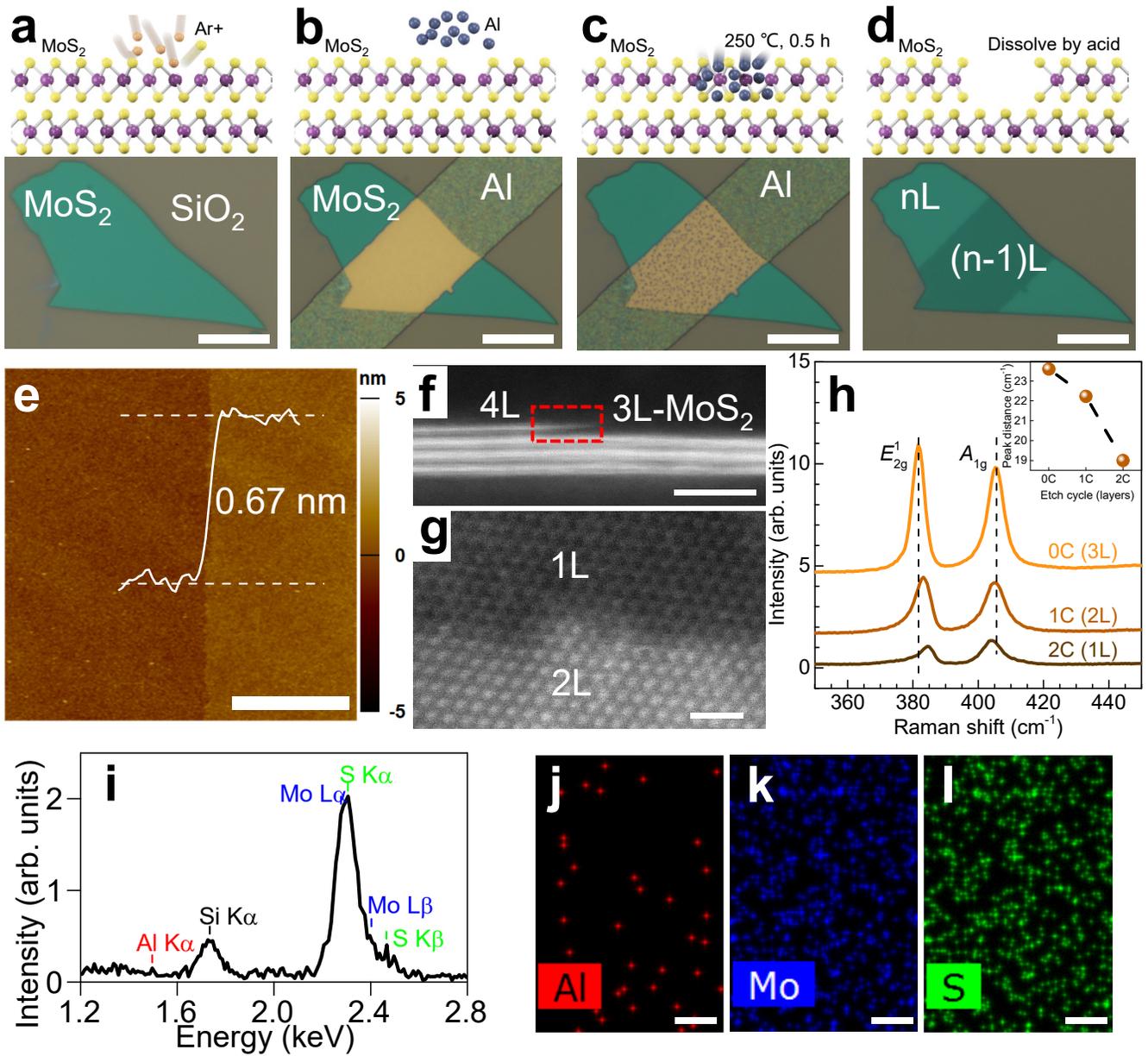

Fig.2

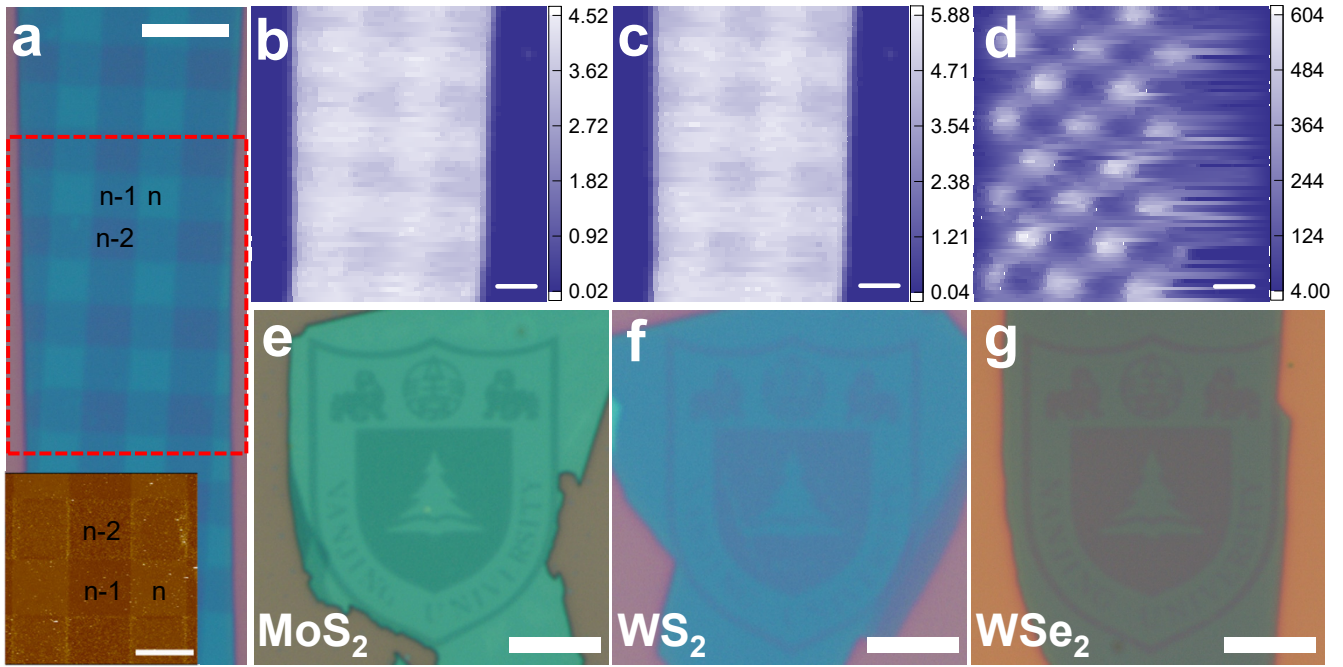

Fig.3

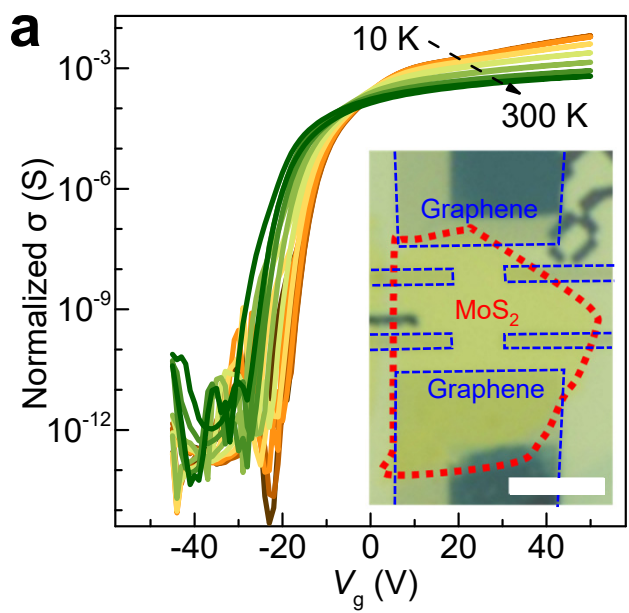
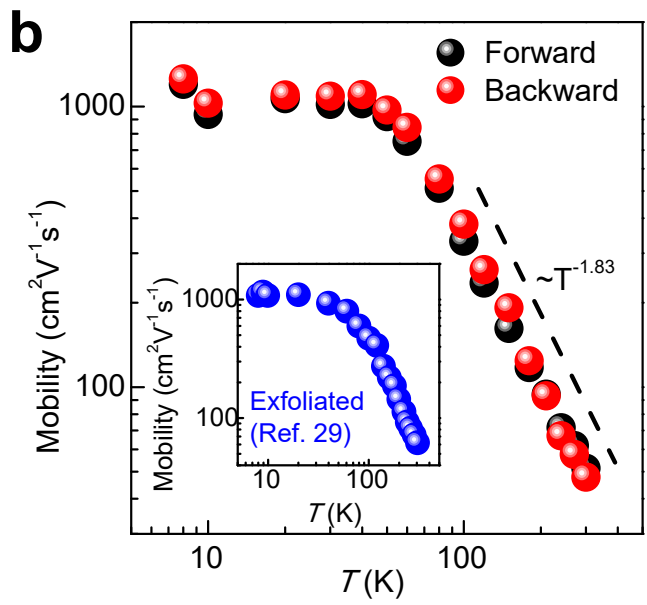
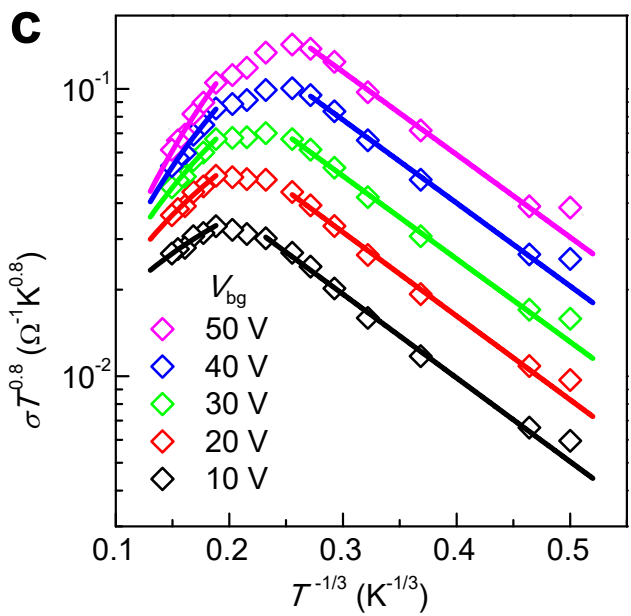
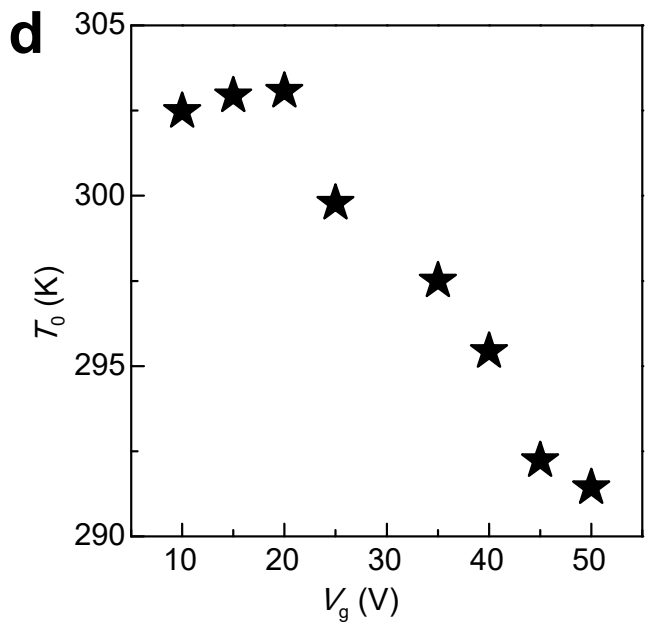

Fig.4

# Non-invasive digital etching of van der Waals semiconductors

Jian Zhou[1,2,†], Chunchen Zhang[1,3,†], Li Shi[4], Xiaoqing Chen[1,2], Tae-Soo Kim[5], Minseung Gyeon[5], Jian Chen[1,2], Jinlan Wang[4], Linwei Yu[1,2], Xinran Wang[1,2,], Kibum Kang[5], Emanuele Orgiu[6], Paolo Samorì[7,*], Kenji Watanabe[8], Takashi Taniguchi[8], Kazuhito Tsukagoshi[8], Peng Wang[1,3,9,*], Yi Shi[1,2,*], Songlin Li[1,2,*]

[1] National Laboratory of Solid-State Microstructures and Collaborative Innovation Center of Advanced Microstructures, Nanjing University, Nanjing, China
[2] School of Electronic Science and Engineering, Nanjing University, Nanjing, China
[3] College of Engineering and Applied Sciences and Jiangsu Key Laboratory of Artificial Functional Materials, Nanjing University, Nanjing, China
[4] Department of Physics, Southeast University, Nanjing, China
[5] Department of Materials Science and Engineering, Korea Advanced Institute of Science and Technology, Republic of Korea
[6] Institut national de la recherche scientifique, Centre Énergie Matériaux Télécommunications, 1650 Blvd. Lionel-Boulet, J3X 1S2 Varennes, Canada
[7] University of Strasbourg, CNRS, ISIS UMR 7006, 8 allée Gaspard Monge, F-67000 Strasbourg, France
[8] National Institute for Materials Science, Tsukuba, Ibaraki 305-0044, Japan
[9] Department of Physics, University of Warwick, Coventry CV4 7AL, UK
[†] These authors contributed equally: Jian Zhou, and Chunchen Zhang.
[*] Correspondence should be addressed to S.L. (sli@nju.edu.cn) or to P.S. (samori@unistra.fr) or to P.W. (wangpeng@nju.edu.cn) or to Y.S. (yshi@nju.edu.cn).

**Table of Content**

## 1. Preparation of cross-sectional STEM specimens

The cross-sectional STEM specimens were fabricated by a lift-out method using focused ion beam technique (FEI Helios 600i dual-beam system). The Al/MoS$_2$ stacks were firstly prepared on SiO$_2$/Si substrates, followed by appropriate thermal annealing. Supplementary Fig. 1**a, b** shows the optical images for typical samples after annealing at 250 and 300 °C, respectively. Before FIB milling for preparing the cross-sectional slices for STEM imaging, metal Pt was deposited onto the Al/MoS$_2$ stacks to protect samples from ionizing radiation during ion milling. Panel **c** and **d** show the corresponding cross-sectional slices for the overall quintuple-layer Pt/Al/MoS$_2$/SiO$_2$/Si stacks for Panel **a** and **b**, respectively. Each layer can be discerned from enlarged SEM images, as shown in Panel **e** and **f**.

## 2. Uncontrollable diffusion between Al and pristine MoS$_2$

Supplementary Fig. 2 shows an extended (~1 μm wide) cross-sectional HR-STEM image for a typical Al/MoS$_2$ (pristine) stack annealed at 300 °C for 1 h, which provide a chance to understand the mechanism of interlayer diffusion and alloying between the sacrificial Al and pristine MoS$_2$ layers. The density of surface defects on the pristine MoS$_2$ is believed to distribute uniformly within the lattice, since it is a freshly exfoliated sheet with high quality. Hence, the thermal diffusion of Al atoms into the MoS$_2$ lattice is expected to be spatially uniform without showing obvious difference along the entire cross section.

However, a mixed combination of diffused (circled by dotted write lines) and undiffused areas (showing sharp boundaries), with roughly equal distribution probabilities, are observed in the extended HR-STEM image shown in Panel **a**. Such non-uniform thermal diffusion behavior can be attributed to the formation of local defects at initial annealing stage, which facilitates subsequent Al diffusion into the defective local MoS$_2$ lattices. Likely, less defects are believed to be present in the undiffused areas and the relative integrity of the lattices hinders the diffusion of Al.

Panels **b** and **c** show the enlarged HAADF images for the diffused and undiffused areas, respectively. In Panel **b**, 6 layers of MoS$_2$ exhibiting expanded layer spacings can be clearly seen in the alloy area. Most expanded layers (the bottom 4 layers) remain parallel to each other, indicating the interlayer diffusion between Al and MoS$_2$ is quite an order process. At such an annealing condition (300°C, 1h), Al consumes MoS$_2$ up to 6 layers and the length of interlayer diffusion reaches ~10 nm. In Panel **c**, sharp layer boundaries and no expanded MoS$_2$ layers are observed at the Al/MoS$_2$ interface in the HAADF image of the undiffused region, implying that no obvious Al/MoS$_2$ alloy is formed.

**3. Fluctuation of diffusion depth in pristine MoS$_2$**

For the stacks of Al and pristine MoS$_2$ annealed at 300 °C for 1 h, we find that there is variation in diffusion depths ranging from 3 to 6 layers in the alloy areas. Typical cross-sectional HAADF images are shown in Supplementary Fig. 3, where the diffused and delaminated MoS$_2$ layers are indicated by the numbers from 1 to 6, counted from the topmost to the bottom layers. The Diffusion depths are 3L, 4L, 5L, 6L from Panel **a** to **d**, respectively. The fluctuation in diffusion depth of 3-6 layer can be attributed to the overall inhomogeneity caused by multiple imperfect surfacial and crystallographic factors from involved materials, including inhomogeneously condensed absorbates on MoS$_2$ surfaces, non-uniform thermal stress, randomly distributed point and line defects inside MoS$_2$, and even the polycrystalline orientations of Al.

Hence, it would be very difficult to well control the depth of diffusion and thus etching in pristine TMDCs without defect engineering. To improve the controllability in diffusion depth, appropriate defect engineering is necessary.

**4. Estimation of lateral resolution in digital etching**

It is possible to estimate the length of lateral diffusion of Al directly from the profiles of the bird-beak shaped alloy edges. Supplementary Fig. 4 shows the typical profiles of the beak edges with different penetration depths, all formed at annealing condition of 300 °C, 1 h. We plot in Panel **a**, for instance, the critical diffusion terminals at the beak edge (P1, P2, P3) and the directions for Al diffusion. By assuming P0 is the starting point of diffusion, the diffusion velocities are $V_{ab}$ and $V_c$ along the lattice ab- and c-directions, and P1, P2, P3 are also the terminals of lateral diffusion at each layer counted from the topmost 1, 2 and 3, respectively. Indeed, all the terminals comprise the profile lines of the beak edges, which is much clearer in Panels **e** and **f** when the penetration depth is larger. The lateral distances from the terminal of the bottom layer (P3 in case of Panel **a**) to those of the upper etching layers are labelled as $d_n$ where n = 1, 2, 3…, representing the number of upper layers counted from the topmost. For simplicity, we use $d_0$ to label the etching distance in the last layer, that is the lateral distance between the terminal of the bottom layer (P3 in case of Panel **a**) and the starting point of etching (P0). Hence, the overall diffusion distance of Al at each layer can be written as $d_0+d_n$, where $d_0$ is unknown but $d_n$ can be accurately measured from the STEM images.

**Supplementary Table 1.** Values of lateral distance measured from the terminal of the bottom layer to those of the upper etching layers.

| Panel No. | $d_1$ (nm) | $d_2$ (nm) | $d_3$ (nm) | $d_4$ (nm) | $d_5$ (nm) |
|---|---|---|---|---|---|
| a | 4.2 | 3.0 | - | - | - |
| b | 4.2 | 3.0 | - | - | - |
| c | 5.5 | 4.2 | 3.0 | - | - |
| d | 7.7 | 6.0 | 4.3 | - | - |
| e | 18.9 | 16.9 | 15.4 | 12.9 | 9.5 |
| f | 22.2 | 20.9 | 18.9 | 15.4 | 10.4 |

By taking the selective of the overall diffusion distance, we can finally obtain the individual diffusion distance ($d_{n+1}-d_n$, or $d_n$ for the last layer) for each etching layer, which represents exactly the lateral diffusion distance of Al at the specific layer when Al penetrates a monolayer vertically along c direction. Thus, the lateral resolution at the topmost layer is 1.5±0.3 nm under the annealing condition of 300 °C, 1 h.

**Supplementary Table 2.** Values of lateral selective distance at each layer.

| Panel No. | 1st layer $d_2-d_1$ (nm) | 2nd layer $d_3-d_2$ (nm) | 3rd layer $d_4-d_3$ (nm) | 4th layer $d_5-d_4$ (nm) | 5th layer $d_5$ (nm) |
|---|---|---|---|---|---|
| a | 1.3 | - | - | - | - |
| b | 1.3 | - | - | - | - |
| c | 1.3 | 1.3 | - | - | - |
| d | 1.7 | 1.7 | - | - | - |
| e | 2.0 | 1.5 | 2.5 | 3.5 | 9.5 |
| f | 1.4 | 2.0 | 3.5 | 5.0 | 10.4 |
| **Average** | **1.5** | **1.6** | **3.0** | **4.2** | **9.9** |
| **Error bar** | **0.3** | **0.3** | **0.7** | **1.1** | **0.7** |

It is likely that the issue of lateral diffusion of Al would arise at elevated annealing temperatures (above 300 °C) or in case of deep etching (removal of several layer at one cycle). However, we have managed to lower the annealing temperature to the safe value at 250 °C by using the defect engineering, and we etch only a monolayer at one cycle.

**5. Theoretical estimation of diffusion energy**

To justify the concept of selective etching and to confirm the effect of TMDC surface defects on interlayer diffusion, theoretical calculations on diffusion barriers are carried out for two types of $MoS_2$ lattices: defective and perfect, as shown in

Supplementary Fig. 5. The calculations are based on methods as follows. Density functional theory (DFT) calculations were performed by using the VASP package. The electron-ion interactions were evaluated by using the projector-augmented wave (PAW) pseudopotential. Exchange-correlation interactions were considered in the generalized gradient approximation (GGA) using the Perdew-Burke-Ernzerhof (PBE) and Heyd-Scuseria-Ernzerhof (HSE) methods. The van der Waals interactions were described by using the DFT+D2 scheme. The climbing image nudged elastic band (CI-NEB) method was used to find minimum energy paths.

For simplicity, we only considered the diffusion process of one Al atom. Panel **a** shows the schematic diagram for the ideal thermal diffusion process for an Al atom into a defective $MoS_2$ bilayer with one S vacancy on the surface. The potential curve at different coordinates of the $MoS_2$ lattice is schematically plotted on the right. Panels **b** and **c** show the estimated diffusion barriers for defective and perfect bilayer $MoS_2$ lattices, respectively. Theoretical calculations indicate that the diffusion barrier for the individual Al atom *via* sulfur vacancies into the defective $MoS_2$ lattice is about 90 meV (Panel **b**), corresponding to the movement of Al atom from the position A to C in the schematic potential curve in Panel **a**. The atomic configurations between Al and $MoS_2$ in different reaction states, i.e., initial/transition/final state (IS, TS, FS), are also given in Panel **b**. Panel **c** shows the diffusion potentials of IS and FS for Al into perfect $MoS_2$ bilayers, where a relatively high diffusion barrier of 750 meV is estimated. This value is more than 8 times higher than that in a defective lattice. According to the Fick's second law of diffusion, an 8-fold variation in diffusion barrier corresponds to a huge difference of $10^6$ in thermal diffusion coefficients at 250 °C. Such a huge difference in diffusion coefficient constitutes the rationale behind the selective etching concept.

## 6. Mechanism of thermal diffusion

According to the Fick's second law of diffusion $\frac{\partial C}{\partial t} = \mathbf{\nabla} \cdot (D\mathbf{\nabla}C)$[1], where $C$ is the concentration of Al atoms in alloy, $t$ is the diffusion time of duration and $D$ is diffusion coefficient, the behavior of thermal diffusion of Al can be well described by solving the diffusion equation above. In the diffusion theory, $D$ depends on annealing temperature $T_a$ following the Arrhenius relation $D = D_0 \cdot \exp(-\frac{\Delta H}{k_B T_a})$, where $D_0$ is a coefficient to be determined from experiment, $\Delta H$ is the activation enthalpy of diffusion, and $k_B$ is the Boltzmann constant.

Using a continuum model for van der Waals materials (Supplementary Fig. 6), we can simply estimate the distribution of Al concentration along the diffusion direction. Since the deposited Al film (10 nm) is much thicker than Al/MoS$_2$ alloy region, we further assume the Al source is an infinite reservoir and its concentration keeps constant at the Al/MoS$_2$ interface during the whole diffusion process. By combining the boundary conditions, the Fick's diffusion law for Al distribution into MoS$_2$ can be simplified as $C(x,t) = C_s \cdot \text{erfc}\left(\frac{x}{2\sqrt{Dt}}\right)$, where the $x$ is the diffusion distance along the one-dimensional flux direction (as shown in Panel **a**), $C_s$ is the constant surface concentration at $x = 0$, and erfc( ) is the complementary error function. By substituting the exact expression of $D$, the function of Al concentration can be written as

$$C(x, t, \Delta H, T_a) = C_s \cdot \text{erfc}\left(\frac{x}{2\sqrt{t \cdot D_0 \cdot e^{-\frac{\Delta H}{k_B T_a}}}}\right). \quad (1)$$

By applying the experimental observation that the penetration depth is 6L for defective areas annealed at 300°C for 1h and assuming that $C(6L) = 0.1\, C_s$ (cutoff criterion), we can extract $D_0 = 4.3 \times 10^{-17}\ \text{cm}^2\text{s}^{-1}$.

Accordingly, we estimate that at 250 °C the diffusion coefficients for Al into defective and perfect MoS$_2$ lattices are $5.9 \times 10^{-18}\ \text{cm}^2\text{s}^{-1}$ and $2.6 \times 10^{-24}\ \text{cm}^2\text{s}^{-1}$ respectively, which features a difference of 6 orders in magnitude. Panel **b** shows the normalized concentration of Al as a function of diffusion depth for various samples and annealing conditions, with taking the concentration value $0.1\, C_s$ as the diffusion cutoff (dashed line in Panel **b**). Within such a criterion, the diffusion depth for Al into defective MoS$_2$ is about 4L at the annealing condition of 250 °C, 0.5 h, while it is almost zero into perfect MoS$_2$. The theoretical calculations further justify the proposed concept of selective etching.

## 7. Optimizing pretreatment conditions

In experiment, we found that the ICP power of 30 W (CCP biasing power: idle mode) is almost the lowest parameter to excite stable Ar plasma in our facility and hence we adopted this value as the power parameter. In addition to theoretical calculation, we also cross-over tested the etching results for the surface pretreated MoS$_2$ sheets under different pretreatment conditions with the duration of plasma irradiation (30 W) varying from 20 to 40 s and thermal annealing $T_a$ changing from 150 to 250 °C.

Supplementary Fig. **a–i** shows the etching results that are the optical images for the tested samples before and after etching for the 9 conditions used in total. Panel **j** shows the image contrast of the etched areas with respect to nearby substrate areas, which can be used to scientifically judge the etching completeness. The criterion is set as 0.99, contrast above which standing for a completer etching process without residues.

The tendency is that samples processed under weak conditions show obvious flake remnants (Panels **a**, **b**, **d** and **g**) or light residues (Panels **c**, **e** and **h**), indicating an incomplete etching for a monolayer. In contrast, a monolayer can be completely removed under intensified conditions (Panels **f** and **i**). Overall, the parameters "30 W, 30 s and 250 °C" is concluded as an optimal pretreatment condition. The detailed conditions are listed as below.

1) Generator RF (ICP) power: <u>30 W</u> (low power to minimize plasma density)
2) Biasing RF (CCP) power: <u>0 W</u> (idle mode to minimize plasma energy)
3) Energy: <u>60–70 eV</u> (estimated energy obtained from the RF generator)
4) Flux of Ar+: <u>$4.5 \times 10^{17}$ cm$^{-2}$s$^{-1}$</u>
5) Duration of irradiation: <u>30 s</u>
6) Dose of Ar$^+$: <u>$1.4 \times 10^{17}$ cm$^{-2}$</u>
7) Net dose to remove S atoms: <u>$2.2 \times 10^{14}$ cm$^{-2}$</u> (estimated from XPS)
8) Etching mechanism: <u>physical bombardment</u>

## 8. Etching yield under optimized pretreatment conditions

We performed XPS characterization to estimate the density of removed sulfur atoms after plasma irradiation. We cut three monolayer CVD MoS$_2$ samples from one wafer and exposed them in Ar plasma (ICP power: 30 W, CCP power: 0 W) under different durations from 0, 90, to 180 s. The untreated sample (duration: 0 s) is used as reference and extended durations of 90 s and 180 s are employed to increase the differences among samples.

Supplementary Fig. 8 shows the XPS spectra and corresponding fittings for the three samples, in order to extract the atomic ratios between S and Mo elements. Evidently, the intensity of the S peak located around 227 eV is reduced when the irradiation duration increases from 0 to 180 s. In panel **d**, the extracted S/Mo atomic ratio is plotted versus irradiation duration. The ratios are 1.96, 1.42 and 1.20 for samples under irradiations of 0, 90 and 180 s, respectively, indicating that the loss of S atoms is roughly proportional to plasma duration. A slight saturation emerges under high duration, which can be attributed to the reduced probability of physical bombardment between S and $Ar^+$ (i.e., reduced etching yield of S atoms by Ar ions) in the highly defective samples.

By assuming a constant etching yield in the first 90 s, for the 30-s irradiation condition employed in manuscript, we estimated a S/Mo ratio of 1.78, a removal of 9.2% S atoms from the topmost layer, and an effective dose of $2.2\times10^{14}$ $cm^{-2}$ by Ar plasma to remove S atoms. Together with the flux of Ar+ ($4.5\times10^{17}$ $cm^{-2}s^{-1}$, ICP: 30 W, CCP: 0W), we estimated a low etching yield of ~0.16% for removal of S atoms at this weak plasma condition.

## 9. Evidence for sub-monolayer attacking depth of Ar plasma

To check the realistic attacking depth of Ar plasma and whether it only attacks the atoms in the topmost layer, we collected Raman spectra for pristine and irradiated monolayers (Supplementary Fig. 9), which show that the signals from lattice vibration disappear only above intensified conditions (ICP = 40 W, idle CCP, 50 s). Under the condition of 30W (ICP) and 50 s, the Raman signal still survives, indicating such a condition cannot destroy the entire lattice. Thus, under the optimized Ar plasma condition (ICP = 30 W, idle CCP, 30 s), the attacking depth into chaclogenides is within monolayer.

## 10. Solutions for wet etching

In order to verify the universal role of both the acid and basic solutions played for wet etching in the final processing step, we show in Supplementary Fig. 10 the results of post-etching with three various acids (sulfuric acid, phosphoric acid, oxalic acid) and one basic (Tetramethylammonium Hydroxide, TMAH) solutions. It proved that all of the four types of solutions work for the removal of Al and alloy layers. However, it deserves noting that, since the $SiO_2$ substrates can also be etched by strong basic solutions, they should be avoided to use in case of $SiO_2$ as substrates. When the supported substrates are slightly etched, the TMDC layers above the $SiO_2$ substrates would roll up (Panel **d**) or even be rushed away in the basic solution. Hence, basic solutions are compatible only with alkali-proof substrates.

## 11. Preparation of top-view STEM grids

Supplementary Fig. 11 shows typical optical images for as-etched 1L MoS$_2$ (Panel **a**) before and (Panels **b–d**) after transferring onto the STEM grids. Panels **b–d** show the images from low to high magnification ratios. The scale bars are 50, 20, and 10 μm, respectively. In Panel **b**, multiple irregularly shaped MoS$_2$ sheets and a rectangular strip of metal Au can be seen. The Au strip is used as label to mark the location of 1L MoS$_2$ areas to facilitate fast locating during STEM imaging. Panel **c** shows enlarged image covering the Au marker and the MoS$_2$ sheet containing 1L area, while Panel **d** focuses on the 1L MoS$_2$ area that is denoted by dotted lines.

## 12. Statistics on vacancy density of as-etched monolayers

Supplementary Fig. 12**a, b** shows typical high-angle annular dark-field STEM (HAADF-STEM) atomic images for local areas of the as-etched 1L MoS$_2$. In the HAADF imaging mode, the brightest dots and their adjacent slightly less bright ones correspond to the heavy Mo and light S atoms, respectively, as represented in Supplementary Fig. 12**a**. The lattice vacancies, i.e., missing atoms, would appear with a reduced brightness as compared to occupied sites, as indicated by the red arrow in Supplementary Fig. 12**b**. Supplementary Fig. 12**c** plots number of samples versus vacancy density and Gaussian distribution was used to fit the data, which reveals an average value of $1.3 \times 10^{13}$ cm$^{-2}$ with standard deviation of $0.6 \times 10^{13}$ cm$^{-2}$, respectively.

## 13. Estimation of uncertainty for top-view elemental mapping

The uncertainty for EDS elemental mapping for the top-view samples becomes higher than that for the cross-sectional samples because of the nature of 1L thickness and fewer numbers of atoms involved for analysis. EDS signals for various elements including Mo, S, Al, Mg, Ar, Sc, Yb and Gd were carefully collected and some are shown in Supplementary Fig. 13**b–f**. The overall counts collected from detector are summarized and compared in Panel **g**. Among them, the elements Mo and S come from the sample and exhibit the strongest EDS signals in terms of the detector counts over the whole sampling area. The elements Mg and Ar are likely introduced by previous samples and their residues constitute the source of chamber contamination; they show less strong signals. Accordingly, the three selected rear-earth elements Sc, Yb, and Gd, which have never been directly introduced in the chamber and believed to be absent in the chamber, are used for calibrating the noise of imaging detection; they show the weakest signals. The element Al, which is the one to be analyzed and shows signal intensity between Mg and Ar, can have two sources: sample and chamber contamination. In Panel **g**, we careful analyze the levels of chamber contamination and detection noise, as indicated by the blue and green bars, respectively. We estimate that the content of Al residues is within the level of chamber contamination. At the most, the trace of Al residues is about 5% if only ruling out the effect of detection noise.

## 14. Check of Al residues with XPS

We also performed XPS analysis to double check the content of Al residues. The result is quite similar to the EDS analysis above that no noticeable Al content is detected. As shown in Supplementary Fig. 14, the signal from Al 2$p$ peak (around 74.5 eV) is at the level of noise in the XPS spectrum, proving that the Al residue is practically negligible. Note that two 100-nm Au pads were placed to mark the as-etched $MoS_2$ flake and also used as the reference for XPS analysis. The peak at 84.3 eV is from the excitation of the Au 4$f_{7/2}$ level.

## 15. An alternative method for defect engineering

To further verify the crucial role of surface defects played in diffusion enhancement, we also tried other methods in introducing surface defects onto $MoS_2$. Supplementary Fig. 15 show the results for using pre-annealing as the defect engineering. In Panel **a**, an exfoliated $MoS_2$ sheet was pre-treated at 400 °C for 1 h before Al deposition. Then, an Al strip was deposited onto the sheet, followed by thermal diffusion at 250 °C for 0.5 h (Panel **b**) and acid wash (Panel **c**). The result shows that a monolayer of $MoS_2$ was removed successfully and homogeneously, indicating that diffusion is indeed enhanced in defective lattices. Hence, pre-annealing can be used as an alternative method for interfacial defect engineering.

## 16. Universality of the etching method

Besides $MoS_2$, we also applied this method to different TMDC sheets including $WS_2$ and $WSe_2$ to check its universality of the layer-by-layer processing. Supplementary Fig. 16 shows the Raman and photoluminescent spectra for the locally 0C, 1C and 2C processed $WS_2$ and $WSe_2$ trilayer sheets. After such consecutive thinning, local areas with thickness of 1L, 2L and 3L can be obtained accordingly. Panel **a** shows the Raman spectra collected for the processed local $WS_2$ areas whose Raman characteristics resemble those of MoS$_2$; the distance of $E_{2g}^1$ (~355 cm$^{-1}$) and $A_{1g}$ (~418 cm$^{-1}$) modes varies with the number of layers, i.e., etching cycles. The values of peak distance versus etching cycle are summarized in Panel **b**. Panel **c** shows the Raman spectra for the processed local WSe$_2$ areas. Their $E_{2g}^1$ (~251 cm$^{-1}$) and $A_{1g}$ (~249 cm$^{-1}$) modes coincide practically with each other and peak distance cannot distinguish the information of number of layers any more. It was reported that the second-order mode 2LA(M) around 260 cm$^{-1}$ gradually emerges as WSe$_2$ is thinned down from 3L to 1L. Also, the $B_{2g}^1$ peak of WSe$_2$ shows alternative odd-even dependency with thickness decreasing, which is coincided with published result using 488 nm laser[2–4]. The unusual thickness dependence exhibited by the three modes corroborates that the digital layer-by-layer etching technique on TMDCs is applicable.

The PL characteristics of MoS$_2$ and WS$_2$ and WSe$_2$ were also added and listed as Panels **d-f**, respectively, showing a strong PL excitation due to the transition to direct band gap when the flakes are etched down to monolayer. All the three thickness dependent PL characteristics are inconsistent with the exfoliated counterparts (WS$_2$, WSe$_2$:[5]; MoS$_2$:[6].) reported in literature.

### 17. Electrical properties of as-etched MoS$_2$ on SiO$_2$/Si

Systematic electrical characterizations were carried out on the field-effect transistors consisted of as-etched MoS$_2$ fabricated on SiO$_2$/Si substrates, as a comparison with those etched by other methods and, also, as a reference to those supported by ultraclean h-BN substrates. In Supplementary Fig. 17**a, b**, the transfer curves of the 2L and 1L layers reveal average threshold voltage values at 10.5 and 9.1 V, which are translated into positive doping levels of 2.6 and 2.3 × 10$^{12}$ cm$^{-2}$, respectively. The appearance of overall positive rather than negative doping effect, as expected from Al as electron donor, indicate that the adverse doping induced by the Al residues is practically negligible, being much weaker than the impacts from the heavily doped silicon gates and surface gaseous absorbates. In addition, both devices exhibit a high I$_{on}$/I$_{off}$ ratio of $10^8$, suggesting that the semiconducting nature is well preserved in the as-etched layers.

The average two-probe carrier mobility ($\mu$) of the two devices are calculated to be 16 and 31 cm$^2$V$^{-1}$s$^{-1}$, respectively. These values are comparable to these of exfoliated samples with similar thickness values (Supplementary Fig. 17**c**), thus it can be inferred that the acid wash can remove nearly all the Al residues and provide a clean and fresh surface for the as-etched samples. In Supplementary Fig. 17**c**, we also compared the $\mu$ values of as-etched samples treated with various etching techniques, including thermal[7], laser[8,9], and plasma[10–12]. Apparently, by implementing the concept of selective etching, our etching method yields samples with the highest electronic quality among the in-situ etching techniques established so far. In other word, our method is a truly non-invasive surface etching technique that can preserve the electronic quality to the most.

Electrical characterizations at variable temperature from 10 to 300 K were carried out on an as-etched 2L device to evaluate the impact of extra lattice defects on the transport mechanism in $MoS_2$. Supplementary Fig. 17**d** shows its transfer curves at different temperatures. The linear plots of $I_{ds}$ in the inset highlight that the metal-insulator transition point of the device is around 37 V. The presence of metal-insulator transition at high carrier concentration confirms again the high crystallinity is preserved after etching. We also extracted the hopping characteristic temperature $T_0$ through the 2D Mott variable-range hopping (VRH) equation. The extracted values of $T_0$ is plotted versus $V_g$ in the inset of Supplementary Fig. 17**e**. $T_0$ ranges from $1.1 \times 10^3$ to $4.9 \times 10^5$ K, comparable to the values reported in exfoliated $MoS_2$ on $SiO_2$/Si substrates[13], which indicates that the extrinsic disorder introduced by Al residues is insignificant.

In Supplementary Fig. 17**f**, we plot $\mu$ versus $T$ to further discern the effect of carrier scattering from Al residues. At high $T$, $\mu$ follows a power law with $T$ ($\mu \propto T^{-\gamma}$ with $\gamma \sim 0.71$) in the log-log plot. At low $T$, $\mu$ becomes saturated at ~70 cm$^2$V$^{-1}$s$^{-1}$. Both the values of $\mu$ and $\gamma$ are much lower than the counterparts supported by ultraclean h-BN, indicating that that charge impurities at the $SiO_2$ surfaces play important role. Hence, ultraclean dielectric interfaces are critical for achieving the intrinsic performance of as-etched TMDCs.

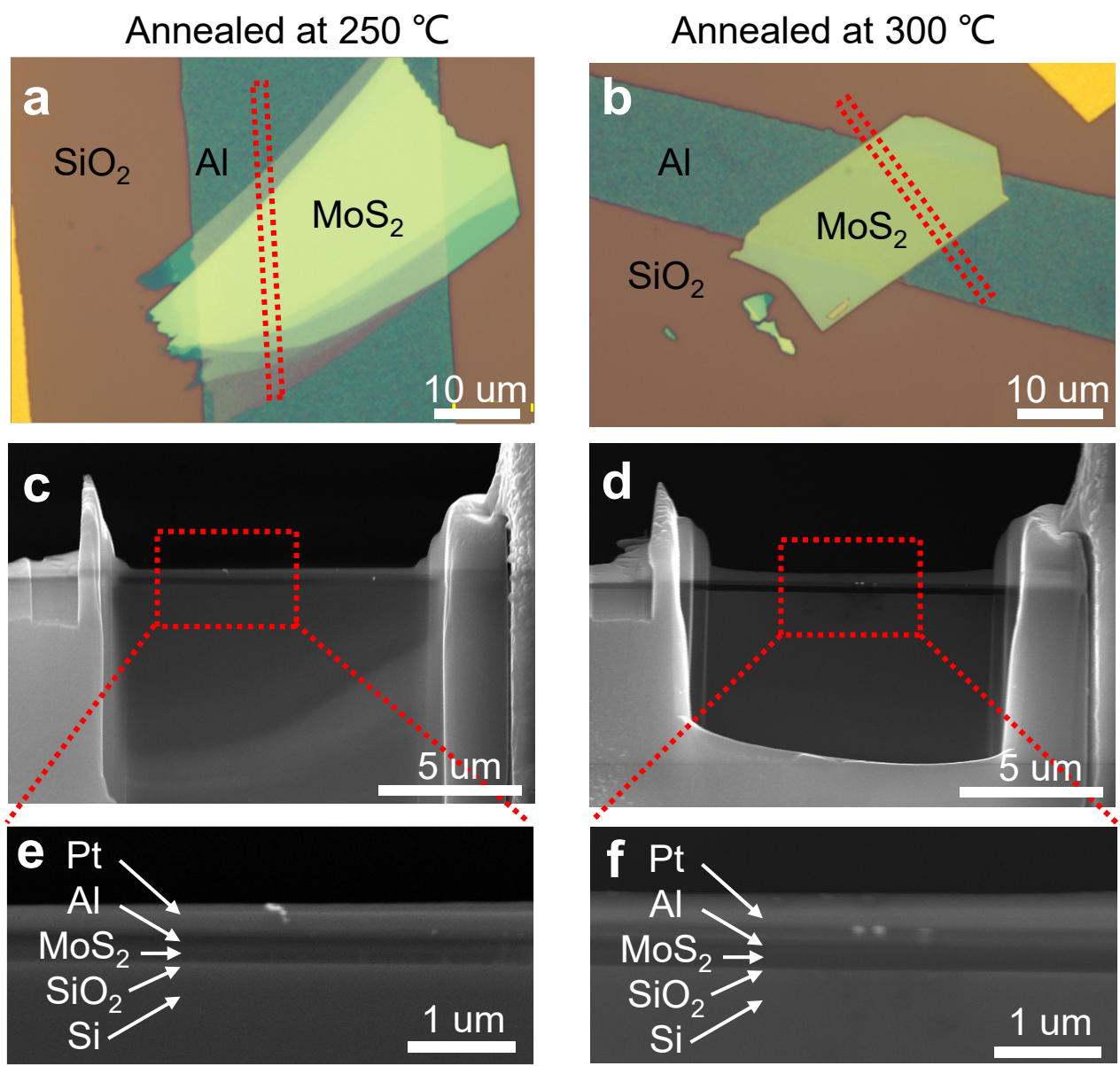

**Supplementary Figure 1. Preparation of cross-sectional STEM specimens by FIB milling. a** and **b,** Typical optical images for Al/MoS$_2$ stacks on SiO$_2$/Si substrates after one-hour thermal annealing at 250 and 300 °C, respectively. **c** and **d,** Corresponding SEM images for the cross-sectional slices after FIB milling. **e** and **f,** Enlarged images for selected areas shown in **c** and **d.** Note that Pt capping layers were deposited before FIB milling to protect samples from local deformation and ionizing radiation.

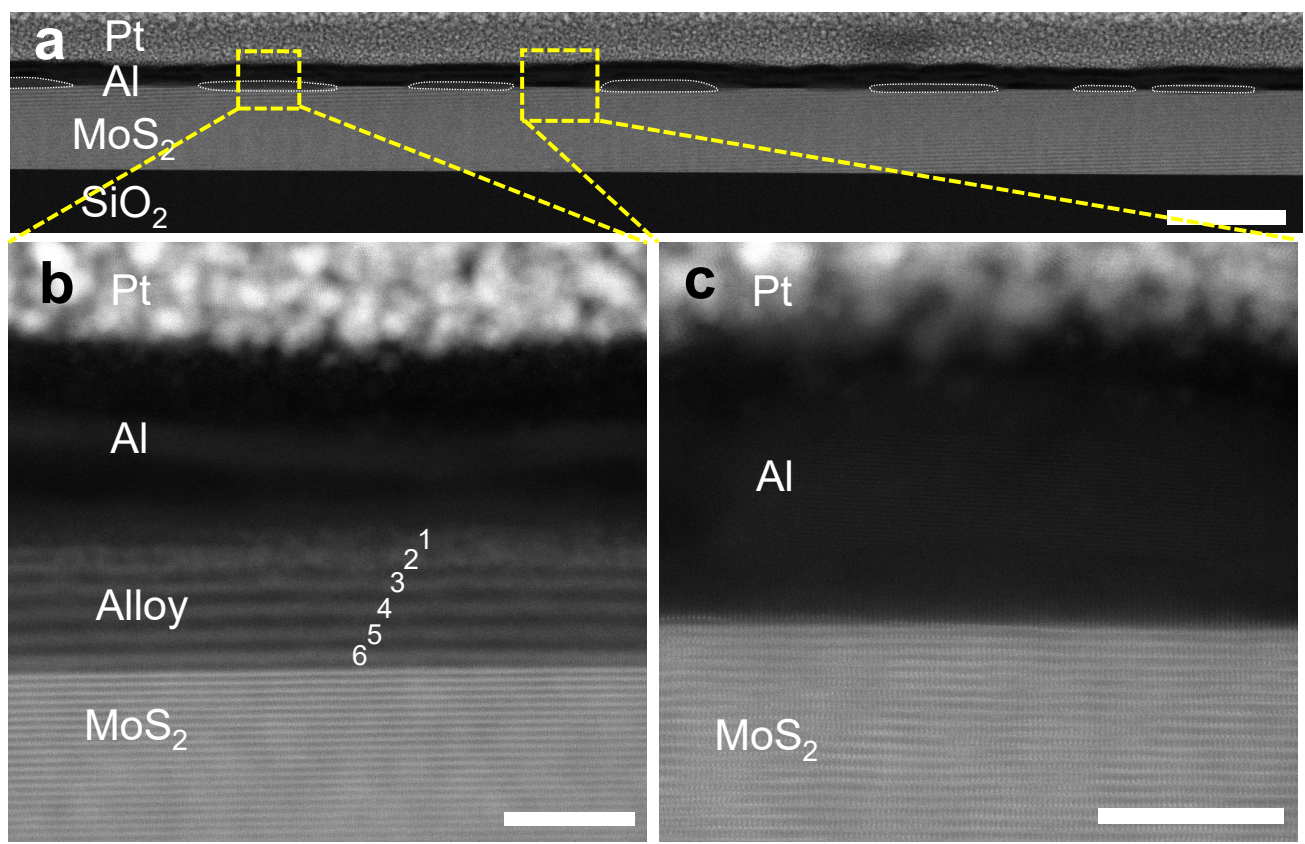

**Supplementary Figure 2. High-resolution cross-sectional STEM images. a,** Large scale HAADF image for a stack of Al and pristine MoS$_2$ annealed at 300 °C for 1 h. Note that the Pt capping layer is employed for sample protection. The alloy regions arising from interlayer diffusion between Al and MoS$_2$ are circled by white dotted lines. The diffused and undiffused areas occur with nearly equivalent probabilities. Scale bar, 100 nm. **b,** Zoom-in HAADF image for the center of an alloy region, where 6 layers of MoS$_2$ are delaminated by Al atoms as indicated by the numbers from 1 to 6. Among the 6 layers, the top 2 layers overlap with each other and become unclear, but the bottom 4 layers still keep the parallel characteristic spatially, which is inherited from their pristine lattice structure. **c,** Zoom-in HAADF image for an undiffused area. No obvious Al/MoS$_2$ interlayer diffusion is observed. Scale bars in panels **b** and **c** are 10 nm.

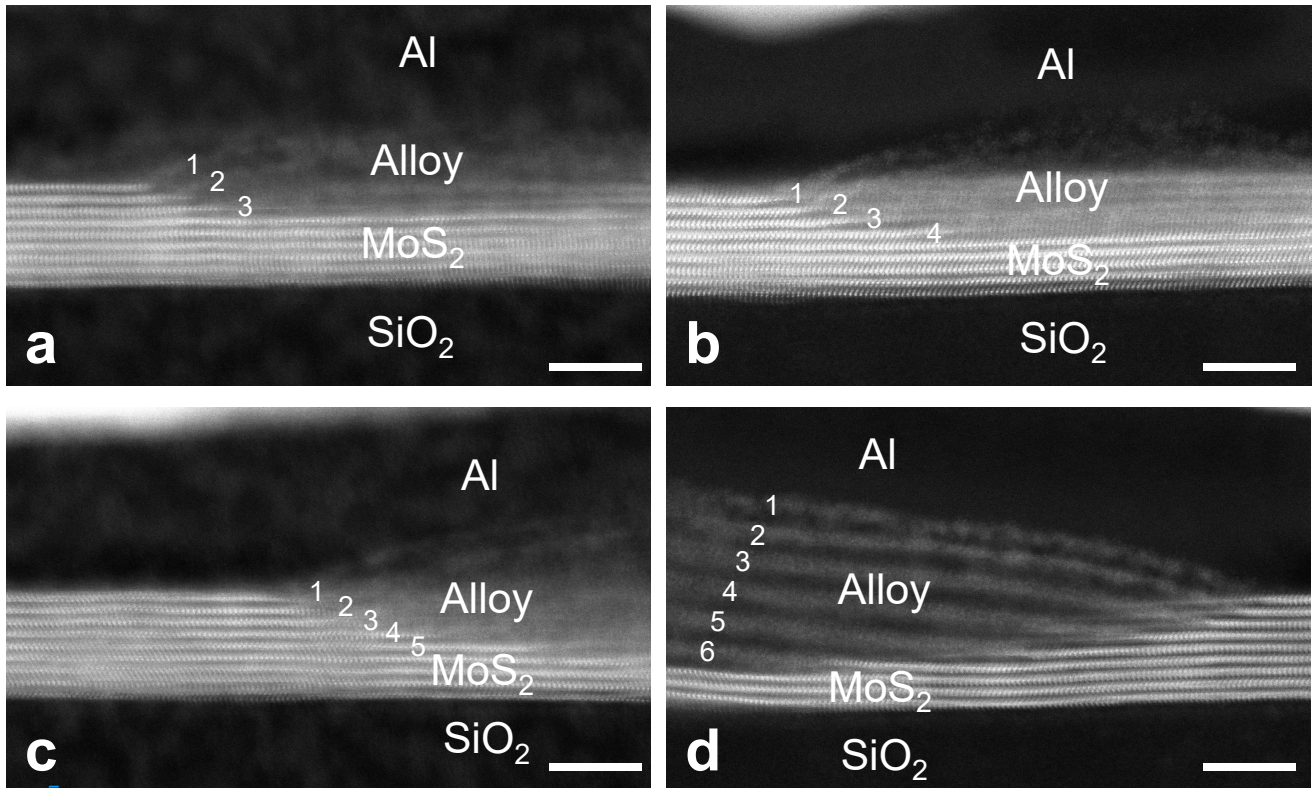

**Supplementary Figure 3. Typical cross-sectional HR-STEM images taken around the beak-shaped alloy edges (droplets) to show the variation in diffusion depth under annealing condition of 300 °C for 1 h.** Diffusion depths are 3L, 4L, 5L, 6L from Panel **a** to **d**, respectively. Scale bar, 5 nm.

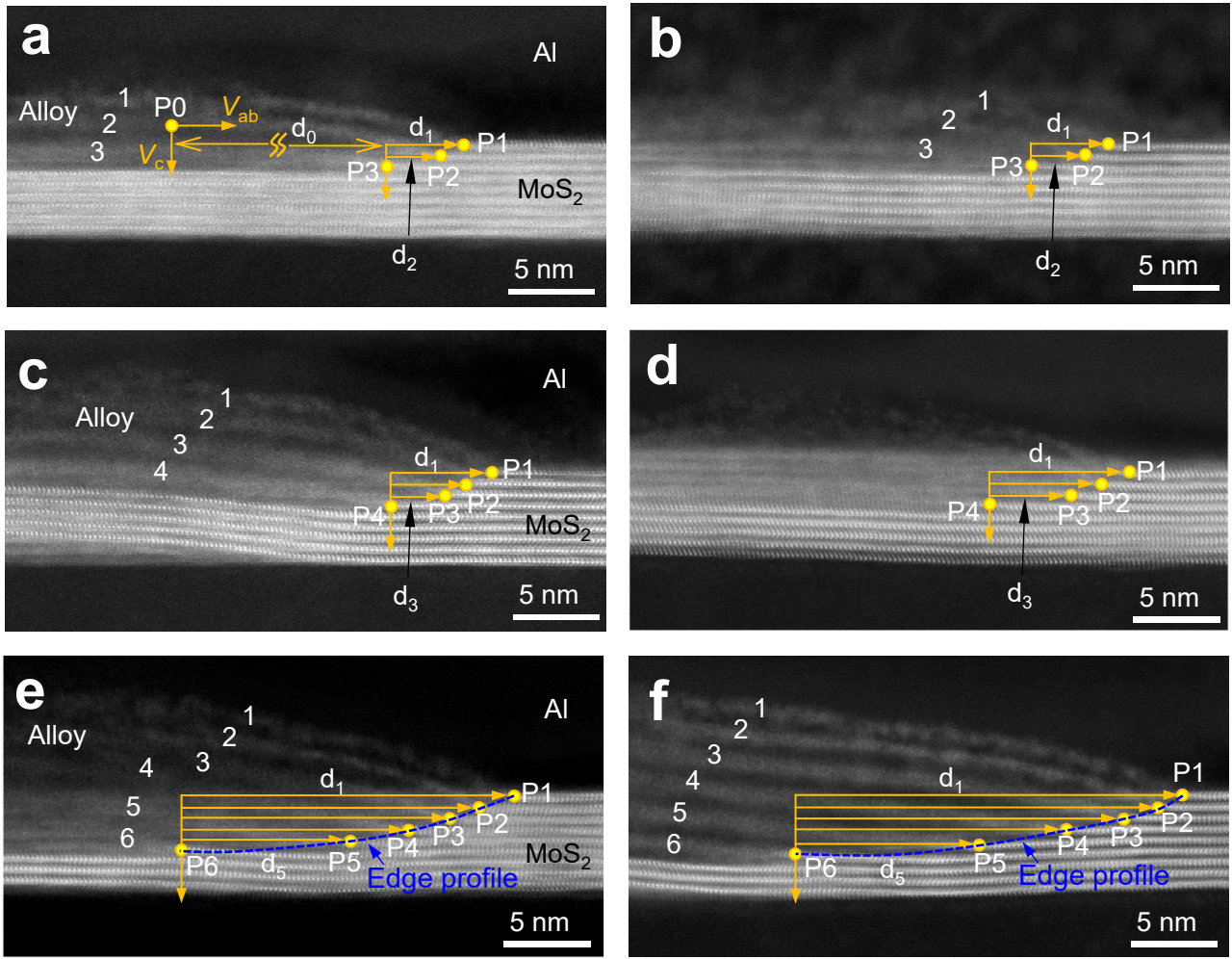

**Supplementary Figure 4. Analysis on the length of lateral diffusion from the profiles of the beak-shaped alloy edges in the Al/MoS$_2$ stacks annealed under condition of 300 °C, 1 h. a-f,** Typical cross-sectional HR-STEM images showing varied diffusion depths of 3L (**a** and **b**), 4L (**c** and **d**) and 6L (**e** and **f**), respectively. The layers of Al/MoS$_2$ alloy are indicated by the number n with n = 1, 2, 3…, representing the numbers of alloy layer counted from the topmost to the bottom. Accordingly, the critical diffusion terminals on the edge profiles are labeled as Pn with n = 1, 2, 3… As an example, the edge profiles in **e** and **f** are labeled by blue dotted lines.

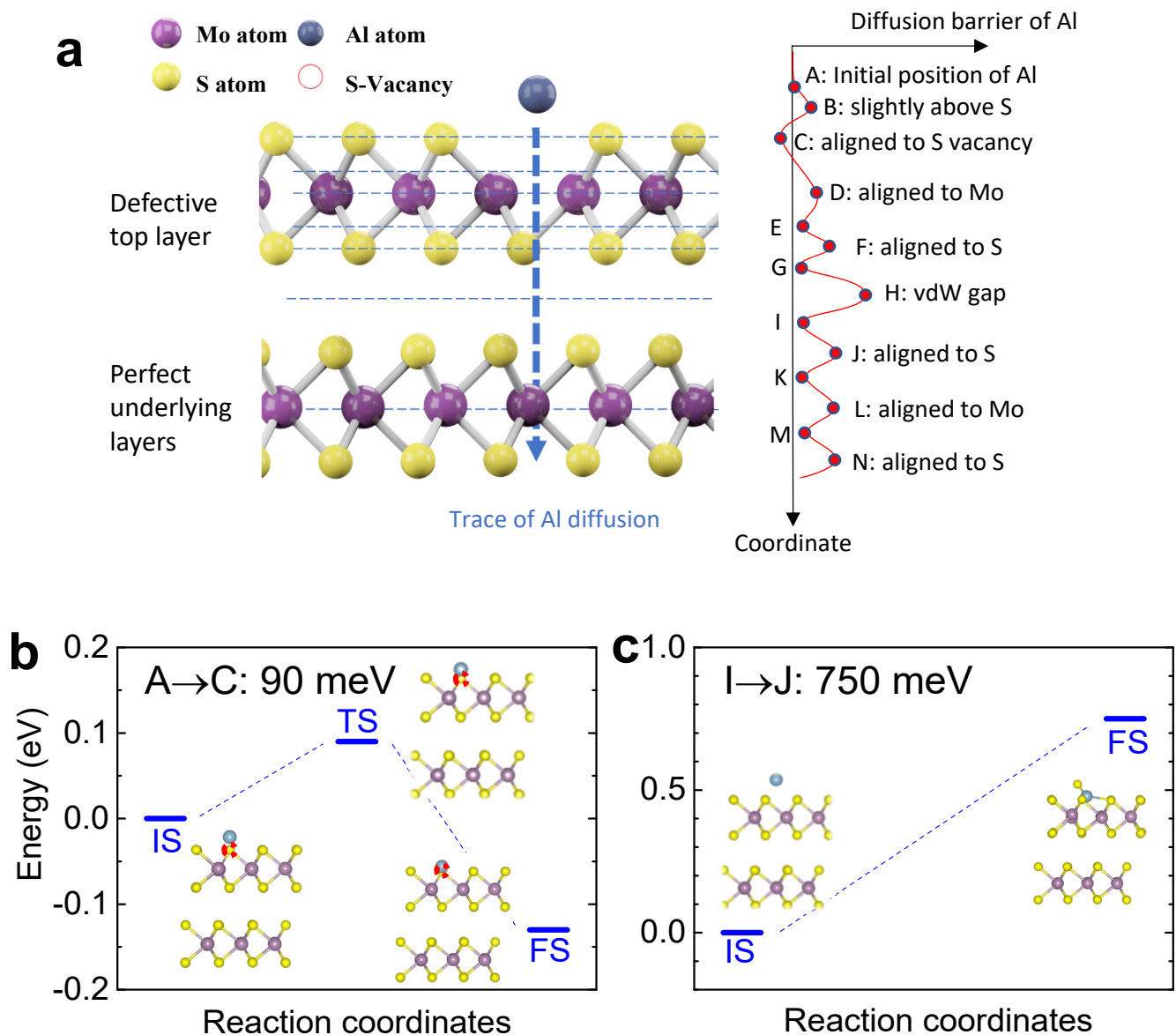

**Supplementary Figure 5. Estimated diffusion energies for Al atoms into defective and perfect MoS₂ lattices. a,** Schematic diagram for an Al atom passing through the defective top and perfect underlying MoS₂ layers, respectively. On the right is schematically plotted the potential curve at different coordinates of the MoS₂ lattice. **b and c**, Diffusion energy curves calculated by density functional theory for Al atoms into the defective (90 meV) and perfect (750 meV) MoS₂ layers, respectively.

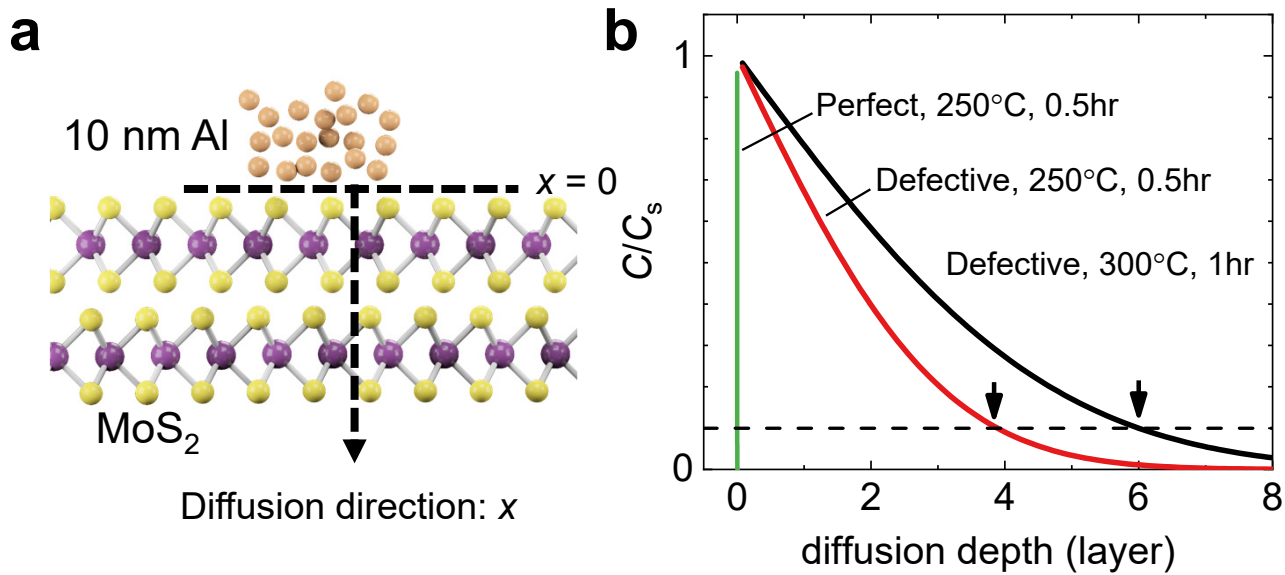

**Supplementary Figure 6. A simplified estimation on the diffusion distribution of Al atoms along the diffusion direction with a continuum model by neglecting the complicated profile of diffusion barrier at different coordinates of the MoS$_2$ lattices. a,** Schematic diagram for the diffusion flux of Al atoms into a MoS$_2$ lattice. The diffusion direction is assigned as $x$. **b,** Calculated normalized concentration of Al element as a function of diffusion depth for three different combinations of sample quality and annealing condition. The criterion of diffusion cutoff is taken as 0.1 $C_s$. Hence, the diffusion depth of Al into defective MoS$_2$ under the annealing condition of 300 °C and 1 h is about 6 layers and is reduced to about 4 layers under the condition of 250 °C and 0.5 h. For a perfect MoS$_2$ lattice, the calculated diffusion depth is practically zero under the condition of 250 °C and 0.5 h.

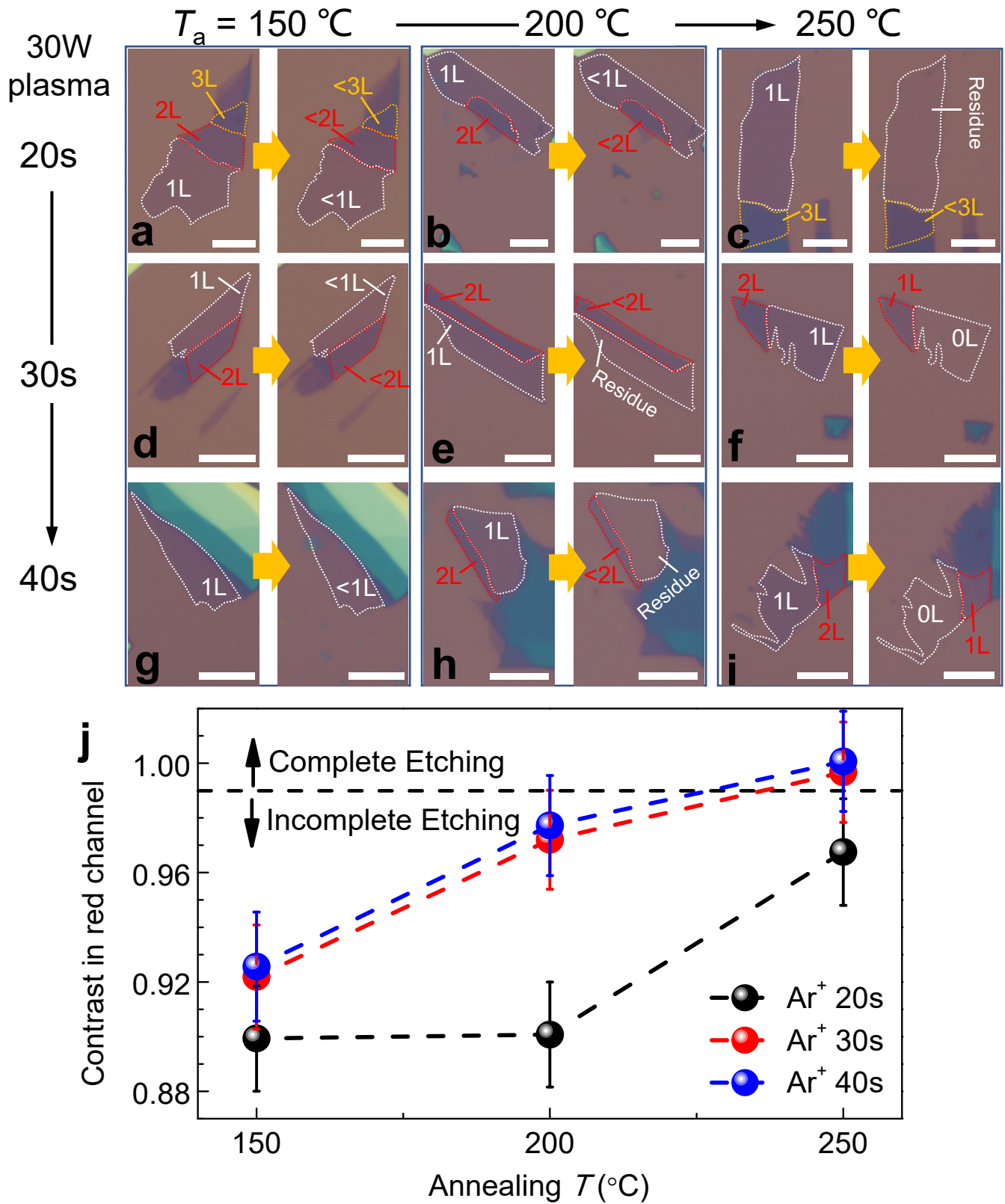

**Supplementary Figure 7. Cross-over test on the etching results under different pretreatment conditions with the duration of plasma irradiation (30 W) varying from 20 to 40 s and thermal annealing $T_a$ changing from 150 to 250 °C. a-i**, Optical images taken before and after etching for checking the residues at varied conditions. Scale bars, 5 μm. **j**, Image contrast of the etched areas with respect to nearby substrate areas, which can be used to scientifically judge the etching completeness. Standard deviations are used as error bars. The criterion is set as 0.99 in contrast for a completer etching process without invisible residues.

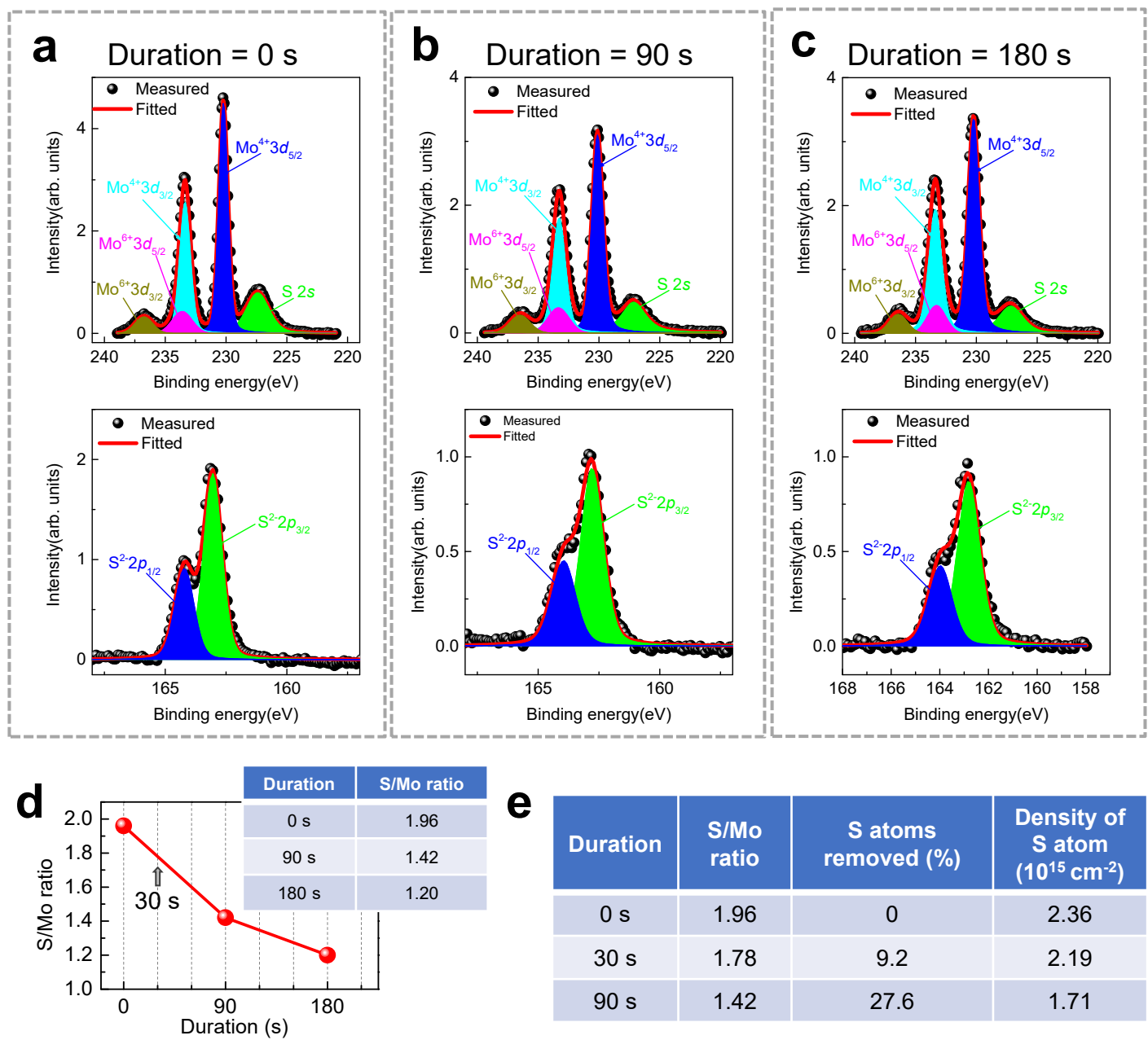

**Supplementary Figure 8.** XPS spectra and fitting results for S/Mo ratio in samples with varied durations of Ar plasma irradiation.

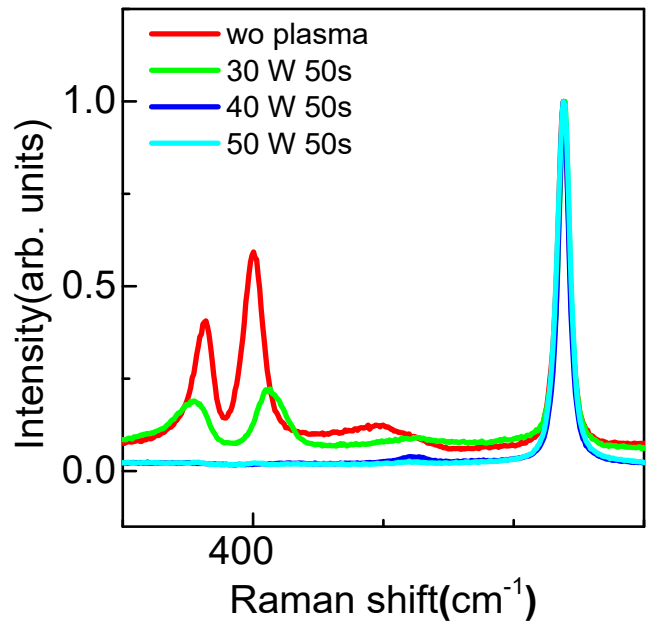

**Supplementary Figure 9.** Raman spectra for pristine and irradiated monolayer MoS$_2$ under different ICP powers of Ar plasma.

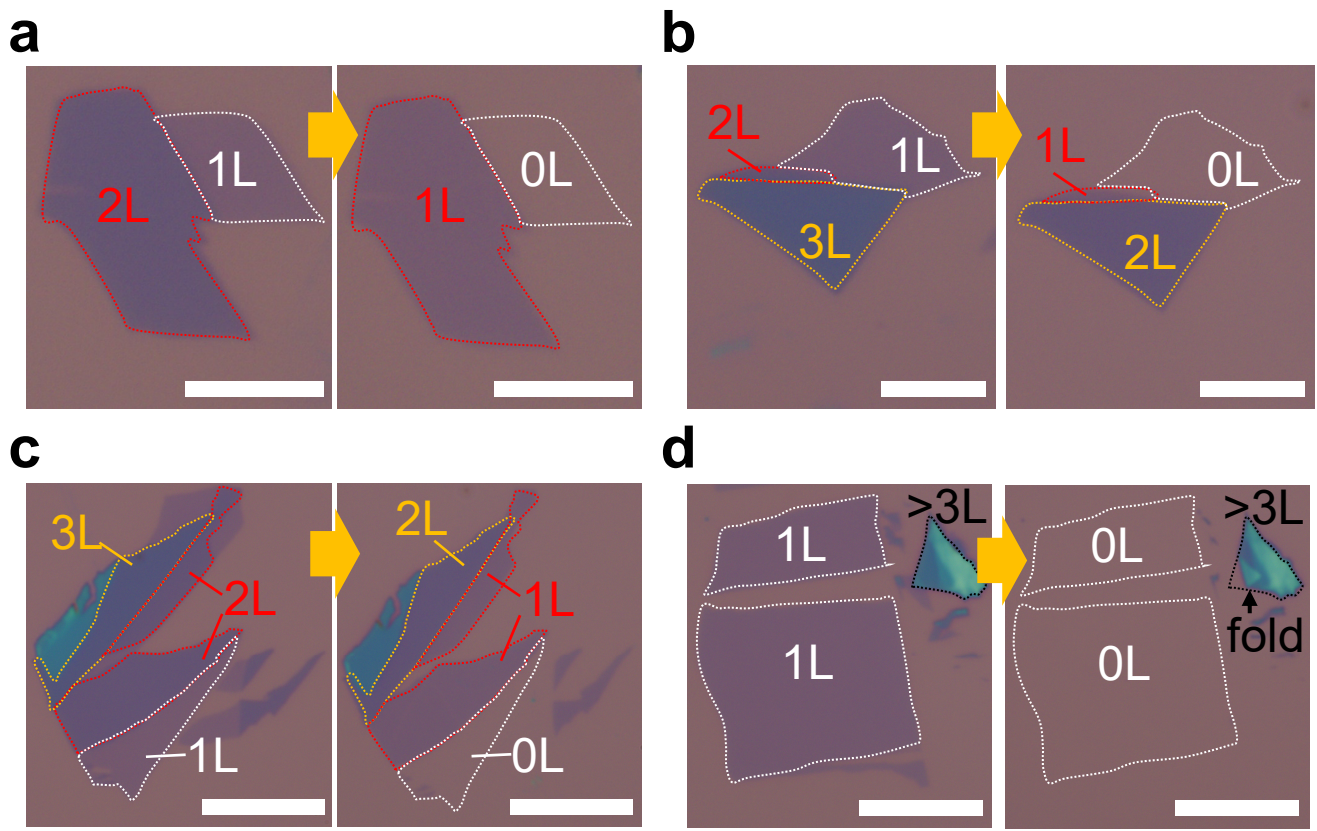

**Supplementary Figure 10. Etching tests by using various acid and basic solutions. a,** sulfuric acid; **b,** phosphoric acid; **c,** oxalic acid; **d,** alkalic TMAH solution. The local $MoS_2$ areas are all labeled by dash line in different colors. Under appropriated pretreatment condition, all tested four solutions above are effective. Scale bar: 5 μm in **a, b**; 10 μm in **c, d**.

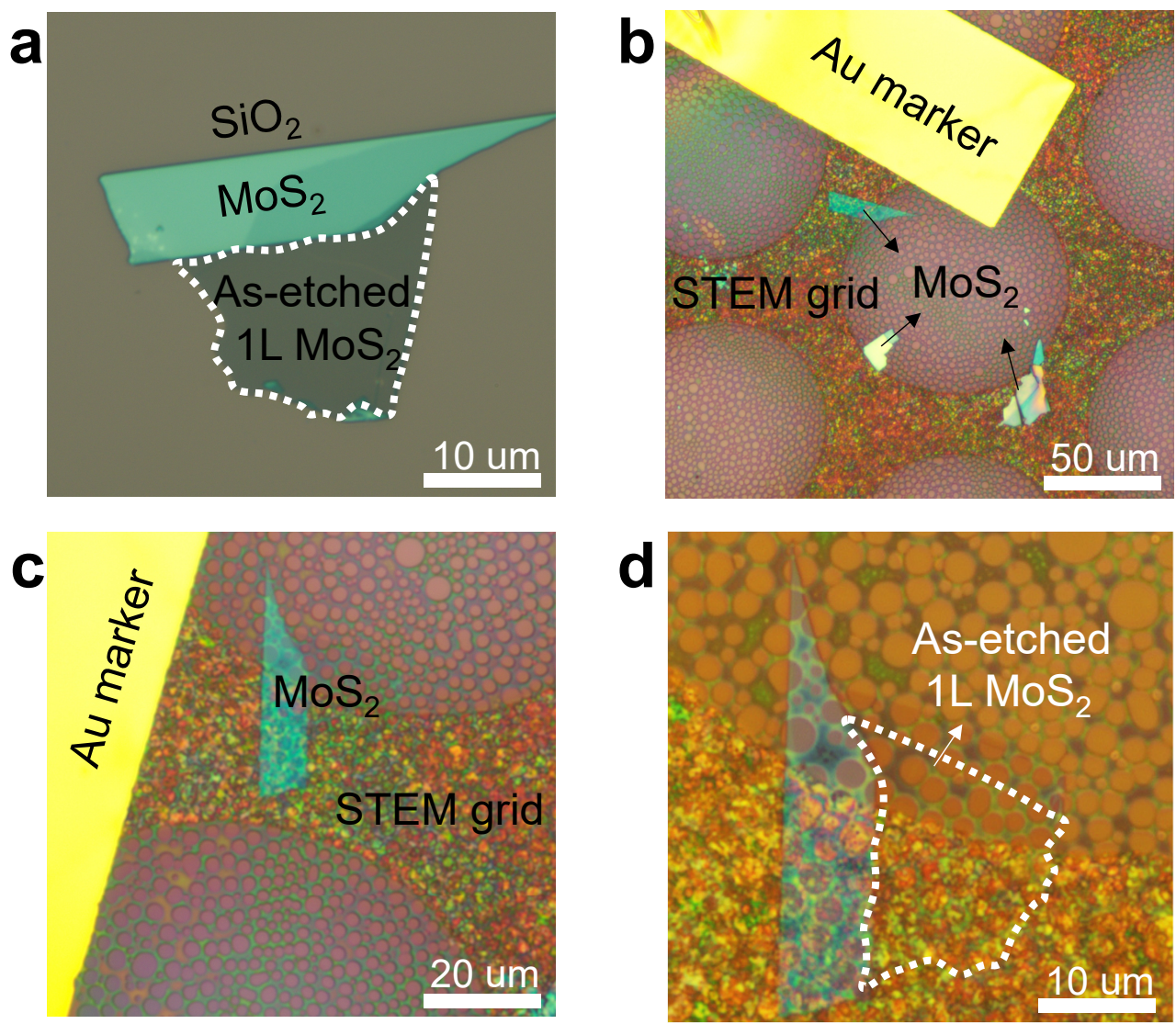

**Supplementary Figure 11. Typical optical images for as-etched 1L MoS$_2$ supported on STEM grids as top-view specimen. a**, As-etched 1L MoS$_2$ on SiO$_2$/Si substrates. **b–d**, Corresponding images for the specimen after transferring onto STEM grid at varied magnification ratios. A gold strip is used as maker to indicate the specimen location.

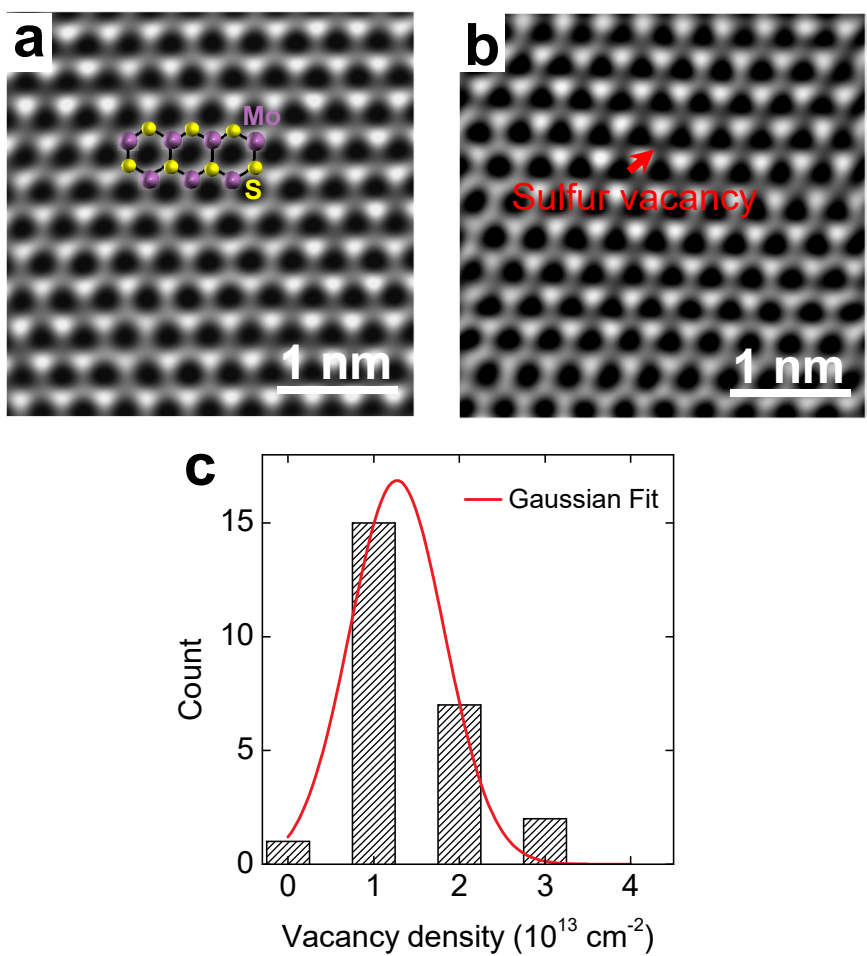

**Supplementary Figure 12. Atomically resolved HR-STEM images for as-etched MoS$_2$ and statistics on vacancy density. a** and **b,** Typical top-view HAADF images for different local areas of an as-etched monolayer, where the most and less bright dots represent the Mo and S atoms, respectively, as labeled by the purple and yellow balls in **a**. A sulfur vacancy can be seen in **b**, as indicated by red arrow. **c**, Histogram and corresponding Gaussian fit on the vacancy density from 25 local areas. The average value and standard deviation of the Gaussian fit are 1.3 and 0.6 ×10$^{13}$ cm$^{-2}$, respectively.

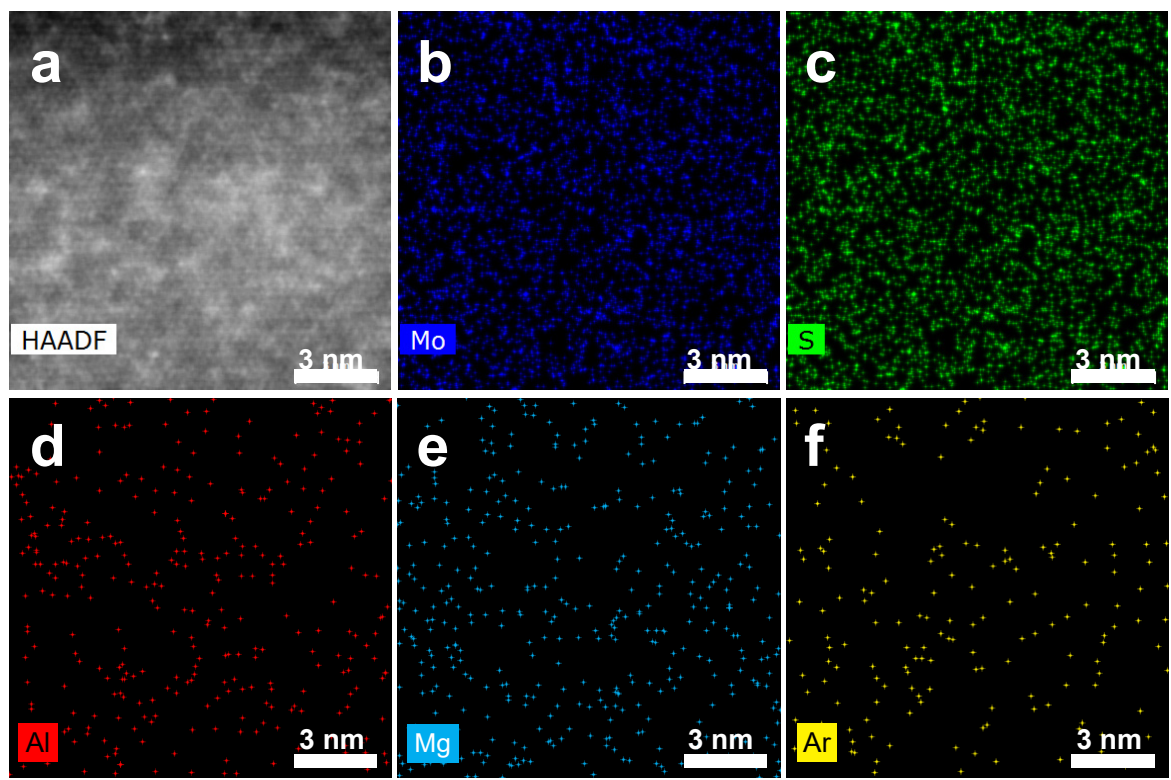
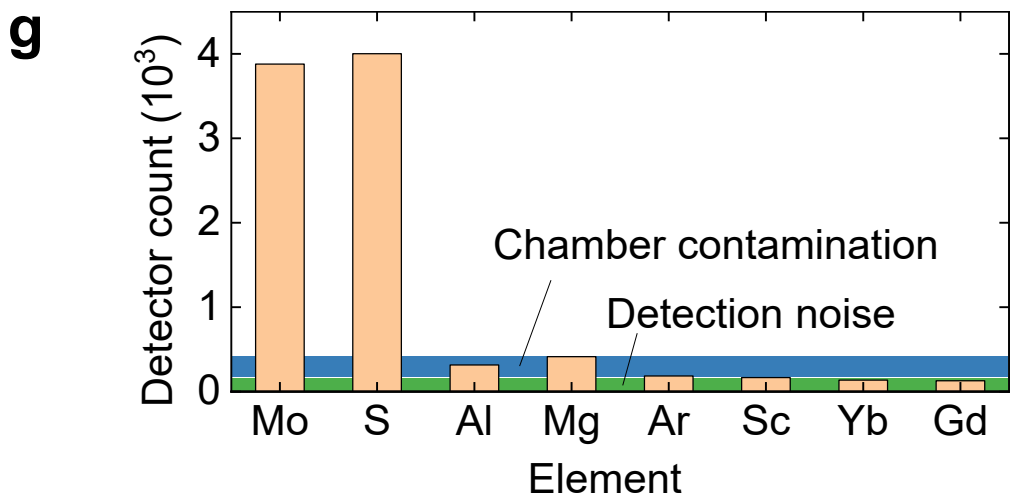

**Supplementary Figure 13. Estimation of uncertainty for the top-view EDS elemental mapping. a,** HAADF image of the area for elemental analyses. **b-f**, EDS mappings for the Mo, S, Al, Mg and Ar elements, respectively. **g,** Statistics for the contents of varied elements. Among them, the detector counts of the rear-earth elements Sc, Yb, and Gd, which have been never directly introduced in the chamber and are believed to be absent in the chamber, are used for calibrating CCD detection noise, and Mg and Ar are likely introduced into chamber from previous samples and hence used for calibrating chamber contamination.

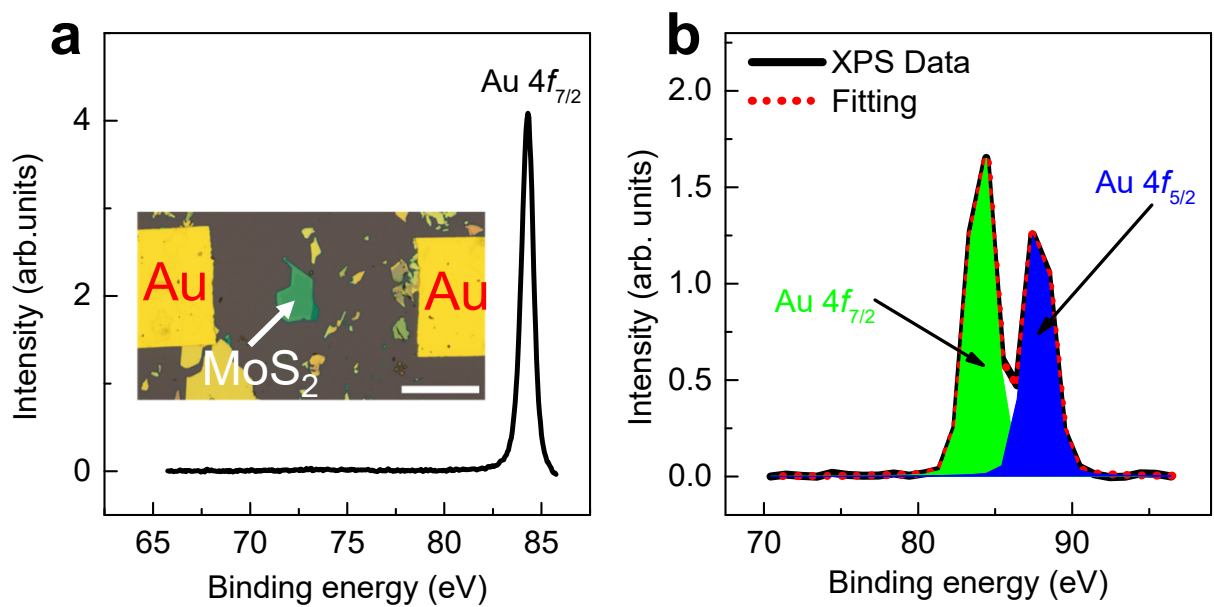

**Supplementary Figure 14. XPS characteristics of as-etched MoS$_2$ for double check on the trace of Al residues a,** XPS spectrum taken around the Al 2p peak where no significant signal was detected from Al residues. The peak at 84.3 eV is assigned to Au 4f$_{7/2}$. Inset: optical image of a MoS$_2$ sample used in XPS. Two Au strips were placed as marker to locate the position of MoS$_2$. Scale bar, 50 μm. **b,** Spectrum with extended energy range in which the doublet from the Au 4f level can be fully observed and the signals of Al remain elusive.

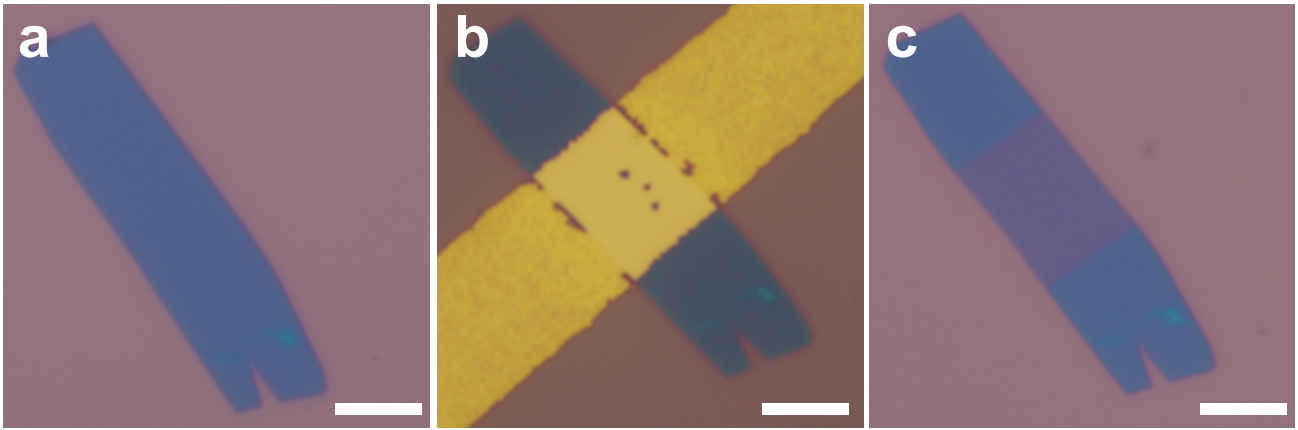

**Supplementary Figure 15. An alternative method for defect engineering via thermal decomposition. a,** Optical image for a typical MoS$_2$ flake under pre-annealing condition of 400 °C, 1 h to form surficial vacancies before Al deposition. **b,** After deposition of an Al strip and thermal annealing at 250 °C for 0.5 h to render Al diffusion. **c,** After acid wash where a monolayer is removed from the center. Scale bar, 5 μm.

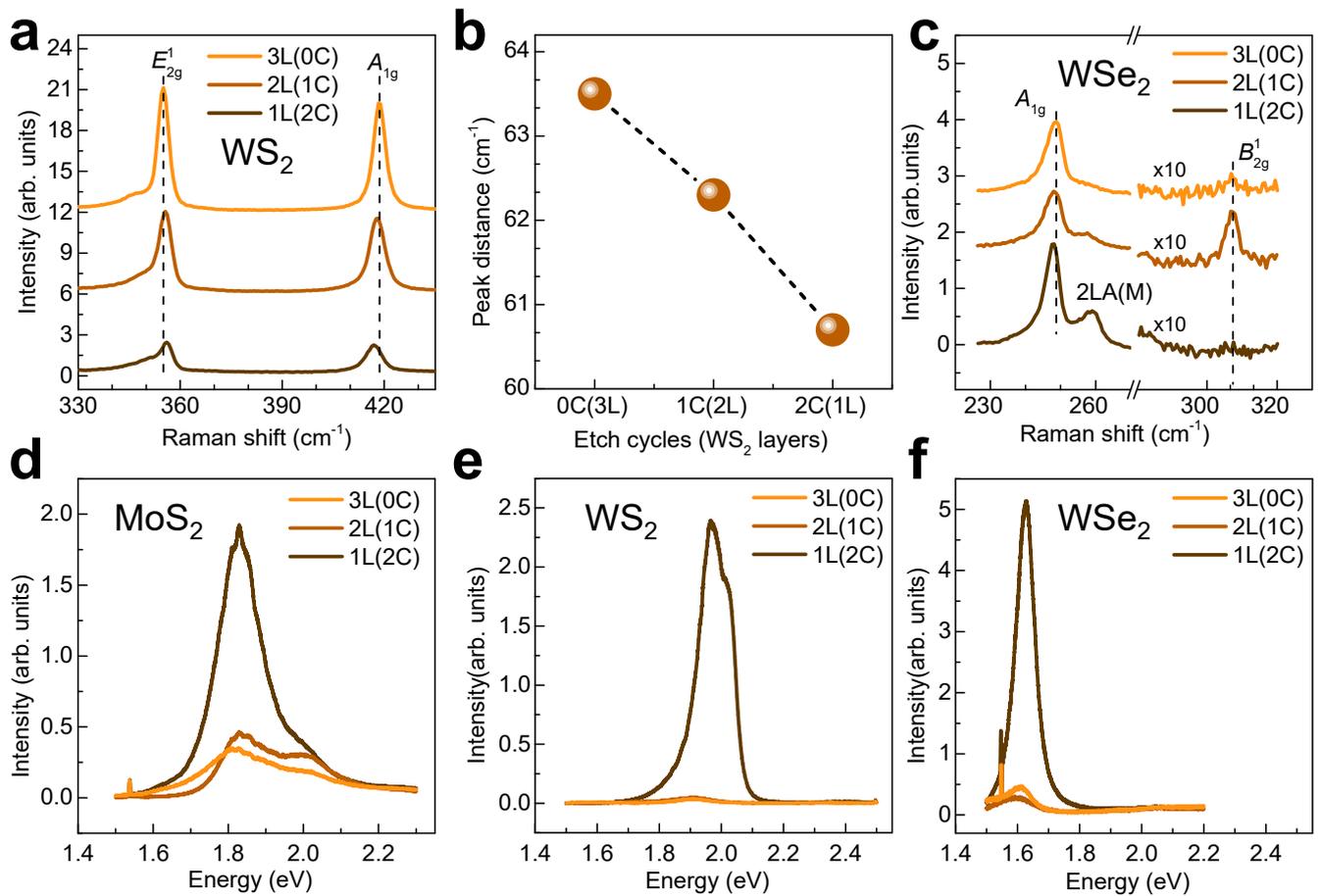

**Supplementary Figure 16. Raman and photoluminescent (PL) spectra taken at different cycles for twice-etched TMDCs, including MoS$_2$, WS$_2$ and WSe$_2$. a,** Raman spectra for local areas of pristine (0C), one-cycle (1C) and two-cycle (2C) etched WS$_2$ (from top to bottom). **b,** Distance between $E_{2g}^1$ and $A_{1g}$ modes versus etching cycles (i.e., number of WS$_2$ layers). The variation of peak distance confirms the applicability of precise layer-by-layer etching on WS$_2$. **c,** Raman spectra for the corresponding 0C, 1C, and 2C etched WSe2 (from top to bottom). The red shift of $A_{1g}$ mode and the emergence of 2LA(M) mode with the decrease in the number of layers suggest the accurate thickness information of 1L, 2L and 3L areas. The $B_{2g}^1$ mode becomes noticeable only in the flakes with odd number of layers, which is consistent with the results reported in literature. **d-f,** Corresponding PL spectra for local areas of 0C, 1C, and 2C etched MoS$_2$, WS$_2$ and WSe$_2$, respectively.

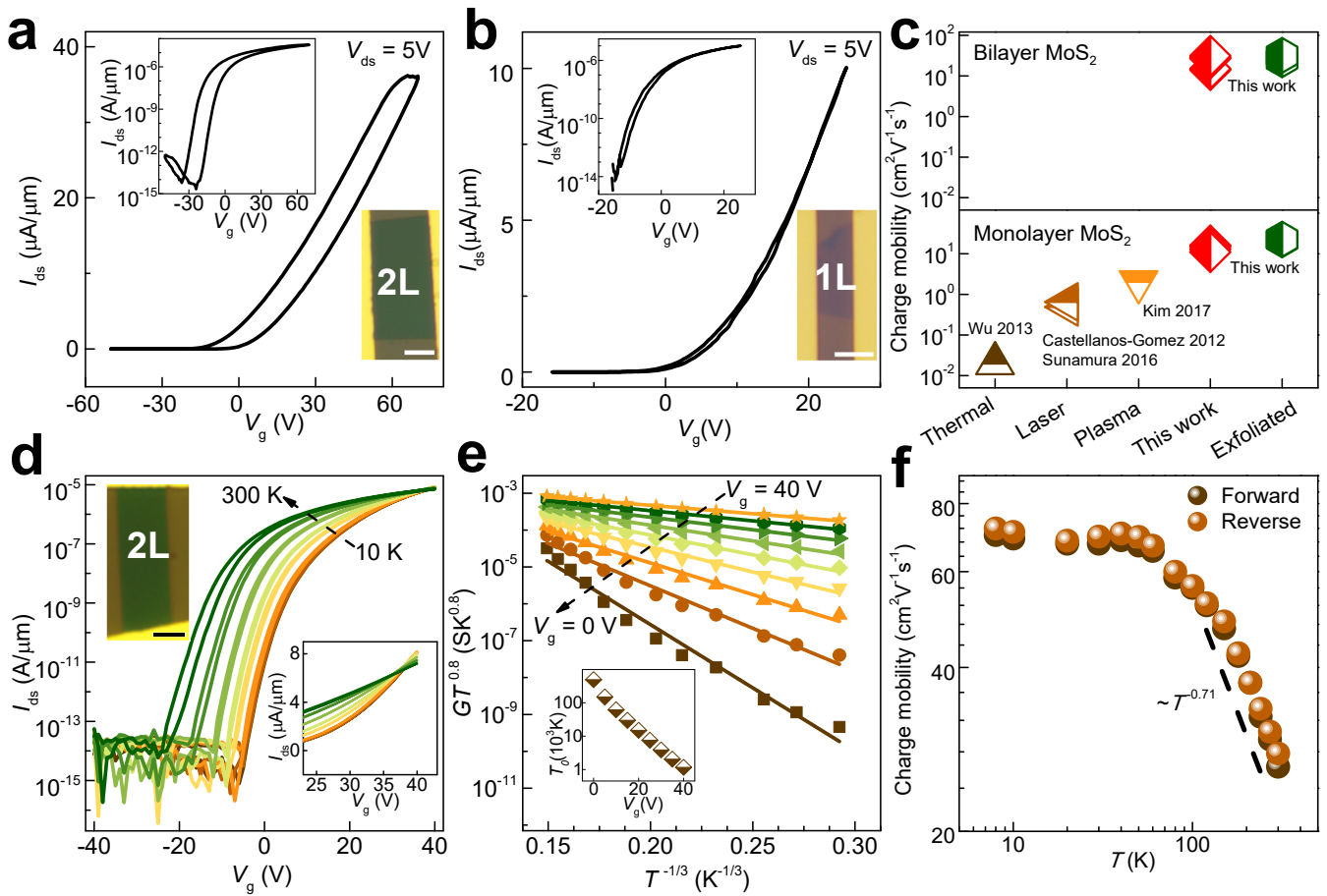

**Supplementary Figure 17. Electrical properties for as-etched MoS$_2$ flakes on SiO$_2$/Si substrates. a-b,** Two-probe transfer characteristics plotted in linear scale for the as-etched bilayer (2L) and monolayer (1L) MoS$_2$. Insets: the corresponding transfer curves in log scale and device images. **c,** Comparison of carrier mobilities of MoS$_2$ sheets prepared by our and other methods (thermal oxidation, plasma, laser and mechanical exfoliation). **d,** Variable temperature measurement of an as-etched 2L MoS$_2$ sheet. Transfer characteristics at different $T$ values of 10 K, 30 K, 50 K, 80 K, 120 K, 180 K, 240 K and 300 K, respectively. The behavior of metal insulator transition (MIT) is highlighted in inset. **e,** Mott VRH model fitting for two-probe conductivity at various gate voltages. Inset: the fitted characteristic temperature $T_0$ versus gate voltage. The values are consistent with those reported in other exfoliated MoS$_2$ devices. **f,** Temperature-dependent field-effect mobility. At $T > 100$ K, the mobility follows μ ∝ $T^{-\gamma}$ with γ = 0.71. The black dashed line (~$T^{-0.71}$) is a guide to the eyes. Scale bar, 4 μm.